\documentclass[apJ]{emulateapj}
\usepackage{textcomp}
\usepackage{epsfig}
\usepackage{amssymb,amsmath}
\newcommand{\apm}{APM~08279+5255}
\newcommand{\pg}{PG~1115+080}
\newcommand{\suzaku}{\emph{Suzaku}}
\newcommand{\xmm}{\emph{XMM-Newton}}
\newcommand{\Rx}{\hbox{X-ray}}

\newcommand{\FeXX}{\hbox{Fe {\sc xx}}}

\newcommand{\FeXXV}{\hbox{Fe {\sc xxv}}}
\newcommand{\FeXXVI}{\hbox{Fe {\sc xxvi}}}
\newcommand{\kms}{\hbox{km~s$^{-1}$}}
\newcommand{\cmsq}{\hbox{cm$^{-2}$}}

\newcommand{\lumin}{\hbox{erg~s$^{-1}$}}

\newcommand{\chandra}{{\sl Chandra}}

\newcommand{\lnh}{\hbox{{\rm log}~$N_{\rm H}$}}

\newcommand{\lip}{\hbox{log~$\xi$}}

\newcommand{\AEn}{\hbox{7--12~keV}}
\newcommand{\OF}{\hbox{observed-frame}}
\newcommand{\RF}{\hbox{rest-frame}}
\newcommand{\Ft}{\hbox{$F$-test}}

\newcommand{\ER}{\hbox{$L/L_{\rm Edd}$}}
\newcommand{\vmin}{\hbox{$v_{\rm min}$}}
\newcommand{\vmax}{\hbox{$v_{\rm max}$}}
\newcommand{\lipR}{\hbox{$2.75\lesssim\lip\lesssim4.0$}}
\newcommand{\LI}{\hbox{low-ionization}}

\begin{document}

\def\sarc{$^{\prime\prime}\!\!.$}

\title{SUZAKU OBSERVATIONS OF NEAR-RELATIVISTIC OUTFLOWS IN THE BAL QUASAR APM~08279+5255.}

\author{C. Saez,\altaffilmark{1} G. Chartas,\altaffilmark{1} and W. N. Brandt\altaffilmark{1}}

\altaffiltext{1}{Department of Astronomy \& Astrophysics,
Pennsylvania State University, University Park, PA 16802,
saez@astro.psu.edu, chartas@astro.psu.edu, niel@astro.psu.edu}

\begin{abstract}
We present results from three \suzaku~observations of the $z =
3.91$ gravitationally lensed broad absorption line quasar \apm. We
detect strong and broad absorption at \RF~energies of $\lesssim$
2~keV (low-energy) and \AEn\ (high-energy). The detection of these
features confirms the results of previous long-exposure (80--90
ks) \chandra\ and \xmm\ observations. The low and high-energy
absorption is detected in both the back-illuminated (BI) and
front-illuminated (FI) \suzaku\ XIS spectra (with an \Ft\
significance of $\gtrsim$99\%). We interpret the low-energy
absorption as arising from a low-ionization absorber with
\lnh$\sim$23 and the high-energy absorption as due to lines
arising from highly ionized (\lipR; where $\xi$ is the ionization
parameter) iron in a near-relativistic outflowing wind. Assuming
this interpretation we find that the velocities in the outflow
range between 0.1$c$ and 0.6$c$. We constrain the angle between
the outflow direction of the X-ray absorber and our line of sight
to be $\lesssim$36$^{\rm o}$. We also detect likely variability of
the absorption lines (at the $\gtrsim$99.9\% and $\gtrsim$98\%
significance levels in the FI and BI spectra, respectively) with a
\RF~time scale of $\sim$1 month. Assuming that the detected
high-energy absorption features arise from \FeXXV, we estimate
that the fraction of the total bolometric energy injected over the
quasar's lifetime into the intergalactic medium in the form of
kinetic energy to be $\gtrsim 10\%$.

\end{abstract}


\section{INTRODUCTION}

Recent observations of nearby galaxies indicate a $M_{\rm
BH}$--$\sigma$ relation \citep[e.g.,][]{Fer00,Geb00}, where
$M_{\rm BH}$ is the mass of the central black hole and $\sigma$ is
the velocity dispersion of the stars in the bulge of the host
galaxy. The presence of a $M_{\rm BH}$--$\sigma$ relation suggests
that a feedback mechanism exists regulating the co-evolution
between the massive black hole at the center of a galaxy and the
formation of its bulge. A possible mechanism of feedback
is quasar outflows. %These winds as stages of gas depletion have
Recent theoretical models demonstrate that quasar feedback can
serve as a fundamental ingredient in structure formation and
galaxy mergers \citep[e.g.,][]{Gra04,Hop05,Spr05}. Quasar outflows
could possibly provide an important source of feedback during the
growth of the super-massive black-holes (SMBHs) in galactic bulges
\citep[e.g.,][]{Fab99}. Another possible mechanism of feedback may
be linked to powerful jets apparently driven by
magnetohydrodynamic forces. As observations indicate, these
powerful jets are predominantly present in radio-loud (RL)
AGNs,\footnote{Radio-quiet (RQ) AGN in general do not contain
large (i.e. kpc) scale collimated jets, although pc-scale jets
have been found in some RQ AGNs \citep[e.g.,][]{Blu96}. Also a
fraction ($\sim$40\%) of radio-quiet AGN could have kpc
radio-structures possibly indicating the presence of an ``aborted
jet'' \citep{Gal06}.} which show a tendency to be found in massive
galaxies and dense environments \citep[e.g.,][]{Bes05}. The
importance of jets as a feedback mechanism has been demonstrated
with recent \chandra\ observations of cavities in clusters of
galaxies and giant elliptical galaxies \citep[e.g.,][and
references therein]{McN07}. The injection of power into the
Intergalactic Medium (IGM) by radio jets is a promising feedback
mechanism that may explain the suppression of cooling flows in the
centers of clusters of galaxies \citep[e.g.,][]{Fab00,McN00,
Sch01,Hei02}. It is not clear, however, if radio jets also
contribute to the feedback process in field galaxies, especially
ones in the redshift range of $z = 1-3 $ where the number density
of galaxy mergers is thought to peak. Most clusters of galaxies
are not formed until $z\approx1$ as inferred from observations
\citep[e.g.,][]{Hil07} and as predicted in theories that consider
a low-density ($\Omega_m \approx 0.3$) Universe
\citep[e.g.,][]{Bah98,You05}. In addition, the fraction of
radio-loud AGNs (RLF) appears to evolve with redshift
\citep[e.g.,][]{Pea86,Sch92,Jia07} and luminosity
\citep[e.g.,][]{Laf94,Jia07}. In particular, the RLF tends to
increase with luminosity and decrease with redshift. For example,
for luminous AGNs ($M_{2500}=-26$; where $M_{2500}$ is the
absolute magnitude at \RF~2500\AA) it is expected that the RLF
declines from 24.3\% to 4.1\% as the redshift increases from 0.5
to 3 \citep{Jia07}. \footnote{As in \cite{Jia07} RLF can be
written in the form of ${\rm log}[RLF/(1 -RLF)]=b_{\rm 0}+b_{\rm
z} {\rm log}(1+z)+b_{\rm M}(M_{2500}+26)$, where $M_{2500}$ is the
absolute magnitude at rest-frame 2500 \AA, $b_{\rm 0}\sim-0.13$,
$b_{\rm z}\sim2.05$, and $b_{\rm M}\sim0.18$.}

Quasar outflows present a promising mechanism of feedback in
high-redshift quasars and possibly in both radio-quiet and
radio-loud AGNs. Powerful winds are observed in Broad Absorption
Line (BAL) quasars, which show deep and broad absorption features
from highly ionized ultraviolet (UV) transitions.  BAL quasars are
also commonly detected to be X-ray weak as a result of high
intrinsic absorption column densities ($N_{\rm H}$) typically in
the range of (1--50)$\times$$10^{22}$ \cmsq\
\citep[e.g.,][]{Gal02, Gal06b}. We note, however, that a recent
survey of BAL quasars obtained from the cross correlation of SDSS
and 2XMM cathalogs by \cite{Giu08} finds no or lower than typical
intrinsic X-ray neutral absorption from that found in optically
selected BAL quasar samples. In the orientation-based BAL model
\citep[e.g.,][]{Wey91} quasar winds exist in most quasars;
however, because of the relatively small opening angles of these
outflows only a fraction of radio-quiet quasars have detectable
BAL features in their UV and/or optical spectra. Models based on
numerical simulations and observations suggest that the winds of
BAL quasars are nearly equatorial
\citep[e.g.,][]{Mur95,Elv00,Pro00}; however, there are a few
observed cases of BAL quasars with outflowing absorbers in the
polar direction \citep[e.g.,][]{Zho06}. Recent studies indicate
that BAL quasars comprise $\sim$15--40\% of the quasar population
\citep[e.g.,][]{Cha00,Hew03,Gib08,Dai08}.

Our current understanding of AGN physics suggests that the most
likely mechanisms to explain the origin of outflows in AGN are
thermal driving, radiation driving (line and continuum), and
magnetic driving. Thermal driving will produce slow winds (with
speeds similar to the sound speed) at large radii ($\sim 10^{4}
R_S$; where $R_S=2GM/c^2$ is the Schwarzchild radius) and with a
relatively small mass-loss rate ($\sim 0.1 M_\odot \rm yr^{-1}$)
\citep[e.g.,][]{Beg83,Kro86}. Therefore, in AGNs thermal driving
is not expected to produce fast and massive winds and consequently
it is likely not a major contributor to feedback.
\newline \indent Given the
typical low temperatures of AGN accretion disks ($T\sim10^5$~K)
and the large gas densities at the base of winds we expect that
initially the absorbing material will have a relatively low
ionization parameter. {For such conditions radiation-driving can
lead to significant acceleration of the absorber.} Magnetic
driving could also be present in strong AGN winds, through the
action of MHD (magnetohydrodynamic) forces \citep[e.g.,][]{Eve05}.
In general, we expect MHD and radiation-pressure forces to act
jointly with the contribution of radiation pressure becoming
increasingly important in sources with higher \ER~
\citep[e.g.,][]{Eve05,Eve07}. Dust in the outflow could also boost
the radiation pressure depending on the spectral energy
distribution (SED) and column density of the material surrounding
the AGN \citep{Lao02,Fab08}. At the moment, evidence for the
presence of near-relativistic outflows in AGN is accumulating
\citep[e.g.,][]{Cha02,Ree03,Pou03,Dad04,Cha07,Zhe08}\footnote{A
recent paper by Vaughan \& Uttley (2008) suggests that some of the
claimed near-relativistic outflows, especially in cases with
narrow absorption lines, are detected at moderate significance
levels and may be spurious. We note, however, that the statistical
significance of the blushifted broad X-ray absorption features
detected in APM 08279+5255 and PG 1115+080 is not disputed.} ;
however, there is no satisfactory model that can produce outflows
with the near-relativistic velocities observed
\citep[e.g.,][]{Mur95,Pro00,Eve05}. We note that recent studies
\citep[e.g.,][]{Che03,Eve05} indicate that with the appropriate
shielding, initial density of the wind, AGN SED and \ER, the
efficiency of the outflows can be significantly increased and the
outflow velocities may approach near-relativistic values.

Due to their high intrinsic absorption, many BAL quasars appear as
faint X-ray sources \citep[e.g.,][]{Gre96, Gal99}. Partly because
of this faintness, it is difficult to detect BALs in X-ray
spectra, and as a consequence, there are only a few cases where
X-ray BALs have been detected in gravitationally lensed BAL
quasars where the magnification effect results in increased
signal-to-noise ratio spectra. Observations in X-rays of the BAL
quasar \apm, the mini-BAL quasar \pg, and perhaps the
low-ionization BAL quasar H~1413$-$117 have suggested the presence
of near-relativistic outflows of X-ray absorbing material in these
objects \citep{Cha02,Cha03,Cha07,Cha07b}. The reported variability
of the high-energy absorption features is over rest-frame
time-scales of 1.8 weeks in APM 08279+5255 (significant detection
of variability) and 6 days in PG 1115+080 (marginal detection of
variability). The analysis of these high-redshift quasars implied
that outflows should have a significant impact in shaping the
evolution of their host galaxies and in regulating the growth of
the central black hole. These observations are particularly
important because they allow us to probe quasar winds at times
close to the peak of the comoving number density of luminous
quasars.
\newline \indent In this paper we describe the analysis of three recent \suzaku~
observations of the lensed BAL quasar \apm. A $\sim$100 ks
observation of \apm\ was performed starting on 2006 October 12
(OBS1), a $\sim$100 ks observation was performed starting on 2006
November 01 (OBS2), and a $\sim$120~ks observation was performed
starting on 2007 March 24 (OBS3).
\newline \indent Unless stated
otherwise, throughout this paper we use CGS units, the errors
listed are at the 1-$\sigma$ level, and we adopt a flat
$\Lambda$-dominated universe with $H_0=70~\kms$~Mpc$^{-1}$,
$\Omega_\Lambda=0.7$, and $\Omega_M=0.3$.

\section{DATA ANALYSIS}

For the reduction and analysis of our observations we used the
\suzaku\ software version~7, which is included in HEASOFT version
6.4. To analyze data from the X-ray Imaging Spectrometer (XIS) and
the Hard X-ray Detector (HXD) we used calibration files that are
part of the \suzaku\ CALDB database released on 2008 April 01.
\footnote{CALDB version 20080401.}

\subsection{XIS data analysis}

\begin{deluxetable*}
{cccccccccc} \tabletypesize{\scriptsize}
%\tabletypesize{\footnootsize}
%\tabletypesize{\small}
\tablecolumns{11} \tablewidth{0pt} \tablecaption{ Log of
observations of \apm. \label{tab:xis}}

\tablehead{
\colhead{Date} & \colhead{OBS ID$^a$} &
\colhead{Telescope} & \colhead{Instrument} & \colhead{Exposure} &
\colhead{Net exp} & \colhead{Net counts$^b$} &
\colhead{$f_{2-10}$$^c$}  }

\startdata

%2006-10-12 & 701057010 & \suzaku & 102.3 ks & 71.3 ks & 2.74$\pm$0.03(24\% bkg) & 3.06$\pm$0.08(28\% bkg) &  4.2$\pm$0.4 & 3.5$\pm$0.5\\
%2006-11-01 & 701057020 & \suzaku & 102.3 ks & 67.9 ks & 2.67$\pm$0.04(26\% bkg) & 2.96$\pm$0.08(29\% bkg) & 3.8$\pm$0.3 & 3.5$\pm$0.4\\
%2007-03-24 & 701057030 & \suzaku & 117.1 ks & 86.4 ks & 2.62$\pm$0.05(25\% bkg) & 3.18$\pm$0.07(28\% bkg) & 4.0$\pm$0.3 & 3.9$\pm$0.3  \\

2002-02-24 & Cha02$^d$  & \chandra & ACIS BI & 88.8 ks & ... & 5723$\pm$76 & 4.3\\
2002-04-28 & Has02$^d$ & \xmm & EPIC pn & 100.2 ks & ... & 12928$\pm$136 &  4.0\\
2006-10-12 & 701057010 & \suzaku & XIS FI & 102.3 ks & 71.3 ks & 7760$\pm$88 &   4.2$\pm$0.4 \\
2006-10-12 & 701057010 & \suzaku & XIS BI & 102.3 ks & 71.3 ks & 3046$\pm$55 &   3.5$\pm$0.5\\
2006-11-01 & 701057020 & \suzaku & XIS FI & 102.3 ks & 67.9 ks & 7121$\pm$84 &  3.8$\pm$0.3 \\
2006-11-01 & 701057020 & \suzaku & XIS BI & 102.3 ks & 67.9 ks &  2855$\pm$78 &  3.5$\pm$0.4\\
2007-03-24 & 701057030 & \suzaku & XIS FI & 117.1 ks & 86.4 ks & 6059$\pm$104 & 4.0$\pm$0.3 \\
2007-03-24 & 701057030 & \suzaku & XIS BI & 117.1 ks & 86.4 ks & 3833$\pm$88 &  3.9$\pm$0.3  \\

\enddata
\tablenotetext{a}{Throughout this paper we identify the \suzaku\
observations 701057010 as OBS1, 701057020 as OBS2, and 701057030
as OBS3.}

\tablenotetext{b}{These counts are obtained in the 0.6--9 keV \OF\
band and in the 0.4--7 keV \OF\ band for the FI and BI chips,
respectively. In each \suzaku\ observation, $\approx$25\% of the
FI counts and $\approx$28\% of the BI counts are background.}

\tablenotetext{c}{The fluxes (in units of $10^{-13}$ergs cm$^{-2}$
s$^{-1}$) in the 2--10 keV \OF\ band are obtained using the
best-fit absorbed power-law model (model 2; \S3) in our \suzaku\
observations. {The fluxes measured in the BI chips are on
average less than those in the FI chips. This is due to
a higher half-power-diameter (HPD) of the XIS1 instrument,
compared to the HPDs of the XIS0, XIS2, and XIS3 instruments}.}

\tablenotetext{d}{In this Table we identify as Cha02 the 88.8 ks
observation of \apm\ performed with \chandra\ in 2002 and analyzed
in detail in \cite{Cha02}. We also identify as Has02 the 100.2 ks
observation of \apm\ performed with \xmm\ in 2002 and analyzed in
detail in \cite{Has02}.}

\end{deluxetable*}

Our data reduction followed the procedures recommended by the
\suzaku\ team for Spaced-Row Charge Injection (SCI) data. The data
reduction was performed on the event files of each XIS instrument
(XIS 0, 1, 2, and 3), and began with recalculating the
PI\footnote{ Each event has a measured ``Pulse Height Amplitude''
(PHA). A calculated ``PHA Invariant'' (PI) value is obtained using
the PHA in combination with the instrumental calibration and gain
drift. For the XIS, the PI column name is ``PI'', which takes
values from 0 to 4095. The PI vs. energy relationship is the
following: $\rm E [eV] = 3.95 \times PI [channel]$.} values of the
unfiltered event files using the XISPI routine. Once the event
files were reprocessed, we used the XSELECT software to apply the
standard screening criteria (see the \suzaku\ ABC
guide\footnote{http://heasarc.gsfc.nasa.gov/docs/suzaku/analysis/abc/})
and obtain ``cleaned'' event files. The data-screening criteria
include selecting events corresponding to ASCA grades 0, 2, 3, 4,
and 6, Earth elevation angles greater than 5$^\circ$ (ELV$>$5),
Earth day-time elevation angles greater than 20$^\circ$
(DYE\_ELV$>$20), exclusion of passages through or close to the
South Atlantic Anomaly (SAA), and cut-off rigidity criteria of
$>$6~$\frac{\rm GeV}{c}$ (COR$>$6). As a final step in screening
the data we removed hot-flickering pixels through the use of the
SISCLEAN routine in XSELECT. The total exposure time of each XIS
chip decreased by $\approx$20\% after the above screening criteria
were applied. Using the clean event files we extracted events in a
circular region centered on the source with a radius of 150 pixels
(2.5$^\prime$). Background events were extracted in an annulus
centered on the source with an inner radius of $\sim$230 pixels
(3.8$^\prime$) and an outer radius of $\sim$430 pixels
(7.1$^\prime$). Our selected background region excludes \apm\ and
the calibration sources located near the corners of the CCDs. The
response matrix files (RMFs) and ancillary response files (ARFs)
were generated using the XISRMFGEN and XISSIMARFGEN routines which
include the correction for the hydrocarbon
contamination\footnote{The XISSIMARFGEN routine incorporates the
XISCONTAMICALC routine which is used to correct the observation
for the XIS optical blocking filter (OBF) contamination. The
absorption due to these contaminants depends on the X-ray energy,
time, detector ID and location on the detector.} on the optical
blocking filter.

For the front-illuminated (FI) XIS chips (XIS 0, 2, 3) we
considered events with energies lying in the range 0.6--10 keV,
while for the back-illuminated (BI) XIS 1 chip we considered
events with energies lying in the range 0.4--8 keV. Due to
calibration uncertainties near the CCD Si K absorption edge at
1.84 keV, events with energies lying in the range 1.7--1.95 keV
were ignored in the analysis of all four XIS chips. In order to
assess systematic uncertainties in the response files, we fitted
the Ni~K$\alpha$ (7.470~keV) calibration line of each instrument.
We found similar positive shifts in the inferred energies of the
calibration lines of each XIS chip ranging from 10 to 20~eV. These
shifts in energy were not large enough to cause any significant
impact on our analysis, and therefore we did not attempt to
correct them. The net source count rate for each XIS chip and each
observation was $\approx$0.04 counts s$^{-1}$, with a background
of $\approx$30 \% of the source rate. The spectra obtained on the
FI chips were combined using the routine ADDSPEC (in HEASOFT
FTOOLS) in order to increase their signal-to-noise ratio. In
Table~\ref{tab:xis} we have included information relevant to the
XIS data analysis. Specifically, this table lists the observation
ID, exposure time, net exposure time (after the screening
process), net counts (for the FI and BI chips) and flux in the
2--10 keV \OF\ (for the FI and BI chips) using the best-fitted
absorbed power-law model (model 2; \S 3). We also have included in
Table~\ref{tab:xis} information from two previous deep X-ray
observations of \apm. These observations correspond to an 88.8~ks
\chandra\  exposure \citep[see][]{Cha02} and to a 100.2~ks
$\xmm$~exposure \citep[see][]{Has02}. The counts collected by the
XIS FI chips for each of our observations are comparable to those
obtained in the \chandra\ observation.

\subsection{HXD data Analysis}

Similarly to the XIS case the clean event files were obtained from
the unfiltered event files following the instructions in the {\sl
Suzaku} ABC guide. The screening criteria are similar to those
applied to the XIS instruments, specifically, we used
ELV$>$5$^\circ$, DYE\_ELV$>$20$^\circ$, exclusion of passages
close to the SAA, and COR$>$6 (units of $\left[\frac{\rm
GeV}{c}\right]$). The HXD-PIN spectrum was extracted from the
cleaned events file described above.  We extracted the source
spectra from the clean files XSELECT. In order to estimate non
X-ray background (NXB) events, we used version 2 of a
time-dependent instrumental background event file (referred to as
the PIN background event file) which was provided by the \suzaku~
team. The PIN background event file was generated with a count
rate that is ten times larger than the real instrumental PIN
background. Therefore, we increased the effective exposure time of
our observed PIN background spectra by a factor of ten. The
exposure time was corrected for dead time using the HXDDTCOR task,
leaving an effective exposure time of $\sim$90\% of the original
exposure time. The effective exposure time of each observation,
together with the count rates (10--40 keV) of the source and NXB
are presented in Table~\ref{tab:hxd}. The NXB does not include the
contribution from cosmic X-ray background (CXB). Therefore the CXB
counts (see Table~\ref{tab:hxd}) have been estimated from a fake
spectrum generated using the FAKEIT command of XSPEC with the
following model \citep[e.g.,][]{Bol87}:

\begin{deluxetable}
{cccccccccc} \tabletypesize{\scriptsize}
%\tabletypesize{\footnootsize}
%\tabletypesize{\small}
\tablecolumns{11} \tablewidth{0pt} \tablecaption{
Log of PIN HXD \suzaku\ observations  of \apm.
\label{tab:hxd}}

\tablehead{ \colhead{Epoch} & \colhead{Net exposure} &
\multicolumn{3}{c}{10--40 keV count rate ($10^{-2}$cts $s^{-1}$)}
\\
& & \colhead{source} & \colhead{NXB} & \colhead{CXB$^a$} }

\startdata

OBS1 & 88 ks & 49.75$\pm$0.24 & 48.17$\pm$0.07 & 2.30$\pm$0.05\\
OBS2 & 89 ks & 50.37$\pm$0.24 & 46.05$\pm$0.07 & 2.39$\pm$0.05\\
OBS3 & 103 ks & 47.07$\pm$0.21 & 44.12$\pm$0.06 & 2.29$\pm$0.05 \\

\tablenotetext{a}{The CXB counts have been estimated from a fake spectrum generated using the FAKEIT command of XSPEC with the model given in equation (\ref{eq:CXB}). }

\enddata

\end{deluxetable}

\begin{equation} \label{eq:CXB}
%\begin{array}{ll}
\frac{CXB(E)}{9.412\times10^{-3}}= e^{-\frac{E}{40{\rm keV}}}
\left(\frac{E}{1{\rm keV}} \right)^{-1.29} {\rm cm^{-2} s^{-1}
sr^{-1} keV^{-1}}.
%\end{array}
\end{equation}

The response file used to fit the PIN spectra was
obtained from the \suzaku\ CALDB calibration files.
The HXD spectral analysis was performed in the 10--40 keV energy
range.

\section{Spectral Analysis}

In this section we fit the \suzaku\ spectra of \apm\ with a
variety of models using the software tool XSPEC version 12. We
also fit the spectra with more realistic models based on the
photoionization code XSTAR. In all spectral models we assume a
Galactic column density of 4.1$\times10^{20}{\rm cm^{-2}}$
\citep{Kar05}. Most of this section concentrates on the analysis
of the XIS spectra; however, in the last paragraph we present
results from the spectral analysis of the PIN spectra of \apm.

\subsection{XIS spectral fits.}
\subsubsection{XSPEC spectral fits.} \label{XSPEC} Each observation of \apm\
provides spectra obtained with the single BI chip (XIS1) and the
FI chips (XIS 0, 2, and 3).  Since the responses of the FI chips
are similar we co-added the FI spectra from each observation. We
note that there is no XIS 2 spectrum of \apm\ for our third epoch
(OBS3) due to the failure of the XIS2 chip.\footnote{On 2006
November 9, about 2/3 of the imaging area of XIS2 became suddenly
unusable
(http://heasarc.gsfc.nasa.gov/docs/suzaku/news/xis2.html).} To fit
the spectra using $\chi^2$ statistics we grouped each XIS spectrum
with a sufficient number of counts. The minimum number of counts
per bin was also chosen to maximize the signal-to-noise ratio in
each bin without losing the features in the spectra and to keep a
similar number of spectral bins in each spectrum ($\approx$70).
The minimum number of counts per bin chosen for the BI chip was 40
for epochs OBS1 and OBS2 and 50 for epoch OBS3. The grouping for
the FI chips was 100 counts per bin for epochs OBS1 and OBS2 and
80 counts per bin for epoch OBS3. Note that for epoch OBS3 we have
increased the binning of the BI spectra due to the longer exposure
and decreased the binning of the FI spectra to compensate for the
loss of XIS2.

\begin{deluxetable*}
{cccccccccc} \tabletypesize{\tiny}
%\tabletypesize{\footnootsize}
%\tabletypesize{\small}
\tablecolumns{18} \tablewidth{0pt} \tablecaption{ Results from
spectral fits to the three \suzaku~observations of \apm.
\label{tab:modn}}

\tablehead{
\multicolumn{2}{c}{} & \multicolumn{3}{c}{FI~~~SPECTRUM$^b$} & & \multicolumn{3}{c}{BI~~~SPECTRUM$^b$} \\
\colhead{Model$^a$} & \colhead{Parameter} & \colhead{Values OBS 1} &
\colhead{Values OBS 2} & \colhead{Values OBS 3} & & \colhead{Values OBS 1}
& \colhead{Values OBS 2} & \colhead{Values OBS 3} }

\startdata

1....... &  $\Gamma$ &  $1.70_{-0.02}^{+0.02}$ & $1.75_{-0.03}^{+0.03}$ & $1.77_{-0.03}^{+0.03}$ & & $1.67_{-0.04}^{+0.04}$ & $1.58_{-0.05}^{+0.05}$ & $1.61_{-0.04}^{+0.04}$\\
&   $\chi^2/\nu$  & 118.3/73  & 125.8/66 & 99.1/71 & & 114.0/71  & 140.8/66 & 100.6/71 \\
&   $P(\chi^2/\nu)$ & $6\times10^{-4}$ &  $1\times10^{-5}$ & 0.02 & & $9\times10^{-4}$ & $2\times10^{-7}$ & 0.01\\
\\

2....... &  $\Gamma$ &  $1.89_{-0.03}^{+0.04}$ & $1.98_{-0.04}^{+0.04}$ & $1.88_{-0.05}^{+0.05}$ & & $1.96_{-0.07}^{+0.07}$ & $1.93_{-0.07}^{+0.07}$ & $1.92_{-0.06}^{+0.06}$ \\
&   \lnh &  $22.83_{-0.10}^{+0.09}$ & $22.92_{-0.10}^{+0.09}$ & $22.66_{-0.20}^{+0.15}$ & & $22.74_{-0.09}^{+0.09}$ & $22.78_{-0.10}^{+0.09}$  & $22.75_{-0.09}^{+0.09}$\\
%&   $\rm f_{2-10}[ergs~cm^{-2} s^{-1}]$  & $4.13~10^{-13}$  & $3.97~10^{-13}$  & $4.12~10^{-13}$  & $3.76~10^{-13}$ \\
&   $\chi^2/\nu$ &  90.5/72  & 87.9/65 & 91.0/70 & & 86.7/70  & 101.3/65 & 59.0/70 \\
&   $P(\chi^2/\nu)$ & 0.07 & 0.03 & 0.05 & & 0.59 &  $3\times10^{-3}$ & 0.82 \\
\\

3....... &  $\Gamma$ &  $1.90_{-0.05}^{+0.04}$ & $1.99_{-0.05}^{+0.05}$ & $1.90_{-0.05}^{+0.05}$ & & $1.97_{-0.07}^{+0.07}$ & $1.94_{-0.08}^{+0.08}$ & $1.94_{-0.07}^{+0.07}$ \\
&   \lnh &  $22.92_{-0.07}^{+0.08}$ & $22.99_{-0.09}^{+0.07}$ & $22.80_{-0.15}^{+0.13}$ & & $22.79_{-0.11}^{+0.11}$ & $22.87_{-0.12}^{+0.16}$  & $22.93_{-0.15}^{+0.08}$\\
&   \lip &  $<$0.18 & $<$0.11 & $<$0.75  & & $<$0.22 & $<$0.68  & $<$0.98 \\
&   $\chi^2/\nu$ &  86.0/71  & 83.2/64 & 87.6/69 & & 84.4/70  & 99.7/64 & 56.6/69 \\
&   $P(\chi^2/\nu)$ & 0.11 & 0.05 & 0.06 & & 0.12 & $3\times10^{-3}$ & 0.86\\
\\

4....... &  $\Gamma$ &  $1.90_{-0.07}^{+0.07}$ & $2.00_{-0.08}^{+0.07}$ & $1.91_{-0.09}^{+0.09
}$ & & $1.96_{-0.09}^{+0.08}$ & $1.92_{-0.08}^{+0.09}$ & $1.91_{-0.06}^{+0.06}$ \\
&   \lnh &  $22.85_{-0.10}^{+0.18}$ & $23.01_{-0.15}^{+0.20}$ & $22.71_{-0.16}^{+0.22}$ & & $2
2.75_{-0.10}^{+0.11}$ & $22.79_{-0.10}^{+0.12}$  & $22.75_{-0.07}^{+0.12}$\\
&   CF & $0.95_{-0.32}^{+0.05}$  & $0.89_{-0.22}^{+0.11}$ & $0.92_{-0.26}^{+0.08}$ & & $0.95_{-0.11}^{+0.05}$ & $0.94_{-0.08}^{+0.06}$  & $0.93_{-0.08}^{+0.07}$\\
&   $\chi^2/\nu$ &  86.9/71  & 85.6/64 & 88.9/69 & & 85.4/70  & 100.9/64 & 57.9/69 \\
&   $P(\chi^2/\nu)$ & 0.10 & 0.04 & 0.05 & & 0.10 & $2\times10^{-3}$ & 0.82\\
\\

5....... &  $\Gamma$  & $1.95_{-0.04}^{+0.04}$  & $1.98_{-0.04}^{+0.04}$ & $1.91_{-0.03}^{+0.05}$ & & $1.95_{-0.06}^{+0.06}$ & $1.91_{-0.07}^{+0.07}$  & $1.92_{-0.06}^{+0.06}$\\
&   \lnh &  $22.95_{-0.07}^{+0.06}$ & $22.98_{-0.08}^{+0.08}$ & $22.82_{-0.13}^{+0.10}$ & & $22.80_{-0.07}^{+0.08}$ & $22.83_{-0.09}^{+0.08}$ &  $22.80_{-0.06}^{+0.07}$  \\
%&  $E_{\rm notch}$[keV]  & $1.95_{-0.03}^{+0.03}$ & $1.99_{-0.05}^{+0.05}$ & $1.91_{-0.04}^{+0.06}$ & & $2.00_{-0.05}^{+0.05}$ & $2.08_{-0.12}^{+0.12}$ & $2.03_{-0.08}^{+0.08}$\\
%&  ${\rm W_{notch}}$[keV]  & $0.81_{-0.07}^{+0.07}$  & $0.79_{-0.10}^{+0.10}$ & $0.82_{-0.08}^{+0.08}$ & & $0.96_{-0.18}^{+0.18}$ & $1.02_{-0.19}^{+0.19}$ & $0.92_{-0.18}^{+0.19}$\\
&  $E_{\rm notch}$[keV]  & $9.59_{-0.15}^{+0.17}$ & $9.75_{-0.24}^{+0.25}$ & $9.40_{-0.20}^{+0.29}$ & & $9.81_{-0.25}^{+0.26}$ & $10.41_{-0.61}^{+0.56}$ & $9.96_{-0.42}^{+0.39}$\\
&  ${\rm W_{notch}}$[keV]  & $3.98_{-0.32}^{+0.35}$  & $3.86_{-0.53}^{+0.47}$ & $4.03_{-0.39}^{+0.44}$ & & $4.71_{-0.93}^{+0.88}$ & $5.01_{-0.96}^{+0.91}$ & $4.52_{-0.88}^{+0.93}$\\
&  ${\rm f_{notch}}$[keV]  & $0.20_{-0.05}^{+0.04}$  &  $0.18_{-0.05}^{+0.05}$ & $0.17_{-0.04}^{+0.05}$ & & $0.26_{-0.05}^{+0.05}$ &  $0.21_{-0.06}^{+0.05}$ & $0.18_{-0.05}^{+0.05}$\\
&  ${\rm EW_{notch}}$[keV]$^c$ &  $0.80 \pm 0.21$ & $0.70 \pm 0.22$ &  $0.69 \pm 0.21$ & & $1.22 \pm 0.33$ & $1.05 \pm 0.36$ & $0.81 \pm 0.28$\\
&   $\chi^2/\nu$ &  62.3/69  & 70.6/62 & 74.2/67 & & 58.1/67 & 90.8/62 & 48.1/67 \\
&   $P(\chi^2/\nu)$ & 0.70 & 0.21 & 0.26 & & 0.77 & 0.01 & 0.96\\
\\

6.......   &$\Gamma$  & $1.93_{-0.04}^{+0.04}$  & $2.00_{-0.04}^{+0.03}$ & $1.93_{-0.05}^{+0.05}$ & & $1.94_{-0.07}^{+0.07}$ & $1.88_{-0.07}^{+0.07}$ & $1.90_{-0.06}^{+0.06}$\\
&   \lnh &  $22.94_{-0.08}^{+0.07}$  & $23.00_{-0.08}^{+0.06}$ &  $22.82_{-0.13}^{+0.11}$ & & $22.80_{-0.08}^{+0.08}$ & $22.81_{-0.09}^{+0.08}$ & $22.79_{-0.08}^{+0.07}$\\
&  $E_{\rm zedge}$[keV] & $7.71_{-0.14}^{+0.15}$  & $8.06_{-0.29}^{+0.17}$ & $7.48_{-0.13}^{+0.12}$ & & $7.47_{-0.13}^{+0.13}$ & $8.32_{-0.33}^{+0.52}$  & $7.74_{-0.23}^{+0.20}$\\
&  ${\rm \tau_{zedge}}$ & $0.32_{-0.07}^{+0.07}$  & $0.30_{-0.08}^{+0.08}$ & $0.31_{-0.08}^{+0.08}$ & & $0.56_{-0.19}^{+0.18}$ & $0.46_{-0.18}^{+0.17}$ & $0.41_{-0.12}^{+0.13}$\\
&  ${\rm EW_{zedge}}$$^c$ &  $1.19\pm0.28$ & $1.12\pm0.28$ & $1.09\pm0.26$ & & $1.86\pm0.59$ & $1.54\pm0.64$ &$1.42\pm0.42$\\
&   $\chi^2/\nu$ &  67.1/70 & 71.1/63 & 74.3/68 & &  64.3/68 & 91.0/63 & 46.1/68  \\
&   $P(\chi^2/\nu)$ & 0.58 & 0.23 & 0.28 & & 0.60 & 0.01 & 0.98 \\
\\

%5.......   &$\Gamma$ &  $1.92_{-0.04}^{+0.05}$  & $2.01_{-0.05}^{+0.04}$ & $1.92_{-0.05}^{+0.05}$ & & $1.93_{-0.07}^{+0.07}$ & $1.91_{-0.07}^{+0.07}$ & $1.93_{-0.06}^{+0.06}$\\
%&   \lnh &  $22.91_{-0.06}^{+0.07}$ & $22.96_{-0.08}^{+0.07}$ & $22.72_{-0.16}^{+0.13}$  & & $22.78_{-0.09}^{+0.09}$  & $22.79_{-0.09}^{+0.08}$ & $22.78_{-0.09}^{+0.08}$ \\
%&  $E_{\rm abs1}$[keV] &  $10.65_{-0.51}^{+0.48}$ & $8.57_{-0.16}^{+0.29}$ &  $7.85_{-0.16}^{+0.16}$ & & $10.45_{-0.34}^{+0.37}$ & $9.36_{-0.23}^{+0.14}$ & $8.32_{-0.38}^{+0.25}$\\
%&  $\sigma_{\rm abs1}$[keV] &  $0.66_{-0.43}^{+0.56}$ & $0.34_{-0.20}^{+0.11}$ & $<$0.34 & & $0.96_{-0.27}^{+0.36}$ & $<$0.65 & $<$0.71 \\
%&  EW$_{\rm abs1}$[keV] &  $0.41_{-0.18}^{+0.24}$ & $0.26_{-0.20}^{+0.15}$ & $0.20_{-0.12}^{+0.21}$ & & $0.98_{-0.39}^{+0.36}$ & $0.38_{-0.14}^{+0.15}$ & $0.37_{-0.15}^{+0.22}$\\
%&   $\chi^2/\nu$ &  75.1/69  & 73.1/62 & 79.9/67 & & 58.9/52 & 65.1/50 & 52.1/67  \\
%\\

7.......  &$\Gamma$ & $1.94_{-0.04}^{+0.04}$  & $2.02_{-0.03}^{+0.03}$ & $1.94_{-0.05}^{+0.05}$ & & $1.95_{-0.06}^{+0.07}$ & $1.91_{-0.06}^{+0.06}$ & $1.92_{-0.06}^{+0.05}$\\
&   \lnh &  $22.94_{-0.08}^{+0.08}$ & $22.99_{-0.04}^{+0.07}$ &  $22.81_{-0.13}^{+0.10}$ & & $22.81_{-0.08}^{+0.08}$  & $22.79_{-0.10}^{+0.10}$ & $22.80_{-0.06}^{+0.07}$  \\
&  $E_{\rm abs1}$[keV] &  $8.10_{-0.12}^{+0.11}$  & $8.62_{-0.18}^{+0.26}$ & $7.83_{-0.09}^{+0.10}$ & & $7.86_{-0.13}^{+0.20}$ & $9.26_{-0.26}^{+0.14}$ & $8.27_{-0.36}^{+0.32}$\\
&  $\sigma_{\rm abs1}$[keV] & $0.29_{-0.17}^{+0.15}$ & $0.37_{-0.18}^{+0.20}$ & $<$0.32 & & $<$0.48 & $<$0.52 & $<$0.74  \\
&  EW$_{\rm abs1}$[keV]$^c$ &  $0.22_{-0.06}^{+0.08}$ &  $0.25_{-0.12}^{+0.10}$ & $0.21_{-0.11}^{+0.09}$ & & $0.23_{-0.10}^{+0.09}$ & $0.34_{-0.20}^{+0.22}$ & $0.39_{-0.12}^{+0.15}$ \\
&  $E_{\rm abs2}$[keV] &  $10.60_{-0.28}^{+0.25}$  & $10.85_{-0.24}^{+0.24}$ & $10.26_{-0.37}^{+0.38}$ &  & $10.53_{-0.30}^{+0.29}$ & $11.32_{-0.80}^{+0.63}$ &$10.98_{-0.48}^{+0.46}$\\
&  $\sigma_{\rm abs2}$[keV]  & $0.78_{-0.22}^{+0.21}$ & $0.36_{-0.16}^{+0.21}$ & $0.72_{-0.33}^{+0.52}$ &  & $0.97_{-0.52}^{+0.54}$ & $<$0.92 & $0.71_{-0.36}^{+0.39}$\\
&  EW$_{\rm abs2}$[keV]$^c$ &  $0.49_{-0.24}^{+0.25}$ & $0.27_{-0.18}^{+0.16}$  & $0.46_{-0.16}^{+0.17}$ & & $1.12_{-0.38}^{+0.36}$ & $<$0.98 & $0.58_{-0.26}^{+0.20}$\\
&   $\chi^2/\nu$ &  58.0/66   & 66.4/59 & 68.8/64 & & 54.0/64 & 79.8/59 & 43.8/64 \\
&   $P(\chi^2/\nu)$ & 0.75 & 0.24 & 0.32 & & 0.81 & 0.04 & 0.97\\
\\

\enddata

\tablenotetext{a}{Model 1 is a power-law with Galactic absorption (PL; XSPEC model wabs*pow);
Model 2 is a power-law with Galactic absorption and intrinsic
absorption (APL; XSPEC model wabs*zwabs*pow);
Model 3 is a power-law with Galactic absorption and ionized-absorption
(IAPL; XSPEC model wabs*absori*pow); Model 4 is power-law with Galactic absorption
and partially covered
absorption (PAPL; XSPEC model wabs*zpcfabs*pow);
 Model 5 is a power-law with Galactic absorption,
intrinsic absorption, and a notch absorber (APL+No; XSPEC model
wabs*zwabs*notch*pow); Model 6 is a power-law with Galactic
absorption, intrinsic absorption, and an absorption edge (APL+Ed;
XSPEC model wabs*zwabs*zedge*pow); Model 7 is a power-law with
Galactic absorption, intrinsic absorption, and two absorption
lines (APL+2AL; XSPEC model $\rm
wabs*zwabs*[pow+zgauss+zgauss]$).}

\tablenotetext{b}{The spectra fitted are the added (ftools
ADDSPEC) spectra of the FI chips (XIS0, XIS2 and XIS3). For OBS 2
the XIS2 CCD was not operational, and only the XIS0 and XIS3
spectra were added for this observation. The BI spectra are taken
with the XIS1 chip.}

\tablenotetext{c}{EW stands for equivalent width, which is defined
as $EW=\int \frac{F_c-F_E}{F_c}dE$, where $F_c$ is the continuum
flux and $F_E$ is the flux in the absorber.}
\end{deluxetable*}

We fit the spectra of \apm\ with the following models: 1)
power-law (PL; XSPEC model wabs*pow) , 2) absorbed power-law (APL;
XSPEC model wabs*zwabs*pow), 3) ionized-absorbed power-law (IAPL;
XSPEC model wabs*absori*pow), 4) partially covered absorbed
power-law (PAPL; XSPEC model wabs*zpcfabs*pow), 5) absorbed
power-law with a notch (APL+No; XSPEC model wabs*zwabs*notch*pow),
6) absorbed power-law with an absorption edge (APL+Ed; XSPEC model
wabs*zwabs*zedge*pow), and 7) absorbed power-law with two
absorption lines (APL+2AL\footnote{We note that if we replace the
APL+2AL model by the XSPEC absorption-line multiplicative model
$\rm wabs*zwabs*gabs*gabs(pow)$, we obtain similar results for the
fitted energies and equivalent widths of the absorption features
found at energies \AEn\ in the \RF. All the results described in
this paper using the APL+2AL can be reproduced using this
multiplicative model.}; XSPEC model $\rm
wabs*zwabs*[pow+zgauss+zgauss]$).
\newline \indent The results of the FI and BI fits with the models described are listed in
Table~\ref{tab:modn}. The error bars of the fitted parameters are
given at the 68\% level ($\Delta \chi^2=1$). For models 2 to 7 we
assume an intrinsic absorber with a redshift of 3.91
\citep{Dow99}. The fits using a power-law model (model~1) are not
acceptable in a statistical sense. We next fit the spectra of
\apm\ with the absorbed power-law model (model~2;
Table~\ref{tab:modn}) assuming an intrinsic absorber. The \Ft\
indicates that fits with model 2 result in a significant
improvement at the $\gtrsim$99\% and $\gtrsim$99.9\% confidence
levels in the FI and BI spectra, respectively, compared to fits
using model~1. Fits with model~2 indicate significant intrinsic
absorption in \apm\ with a column density of \lnh$\approx$23. We
also fit the spectra of \apm\ with more complex models that
included an ionized and partially covered absorber (models 3 and
4), however, these fits did not result in a significant
improvement (\Ft\ significance $<$95\%) compared to the simpler
model 2.

Fits to the spectra of \apm\ with models (models 5--7) that
account for the absorption found between \AEn\ in the \RF\ result
in significant improvements (the $F$-test indicates improvements
at $\gtrsim$99.9\% and $\gtrsim$99\% confidence in the FI and BI
spectra, respectively) compared to fits with models that do not
include this high-energy absorption. We note that the absorption
feature at \AEn\ in the \RF\ corresponds to a significant
detection following the criteria described in \S3 of \cite{Vau08}.
Specifically, we find the ratio of the total equivalent
width\footnote{The equivalent width (EW) is defined as $EW=\int
\frac{F_c-F_E}{F_c}dE$, where $F_c$ is the continuum flux and
$F_E$ is the flux in the absorber.} of the absorption features to
their uncertainty to be $EW/{\sigma_{EW}} \gtrsim3$ in every
observation (see models 5--7 in Table~\ref{tab:modn}).

\begin{figure*}
   \includegraphics[width=14cm]{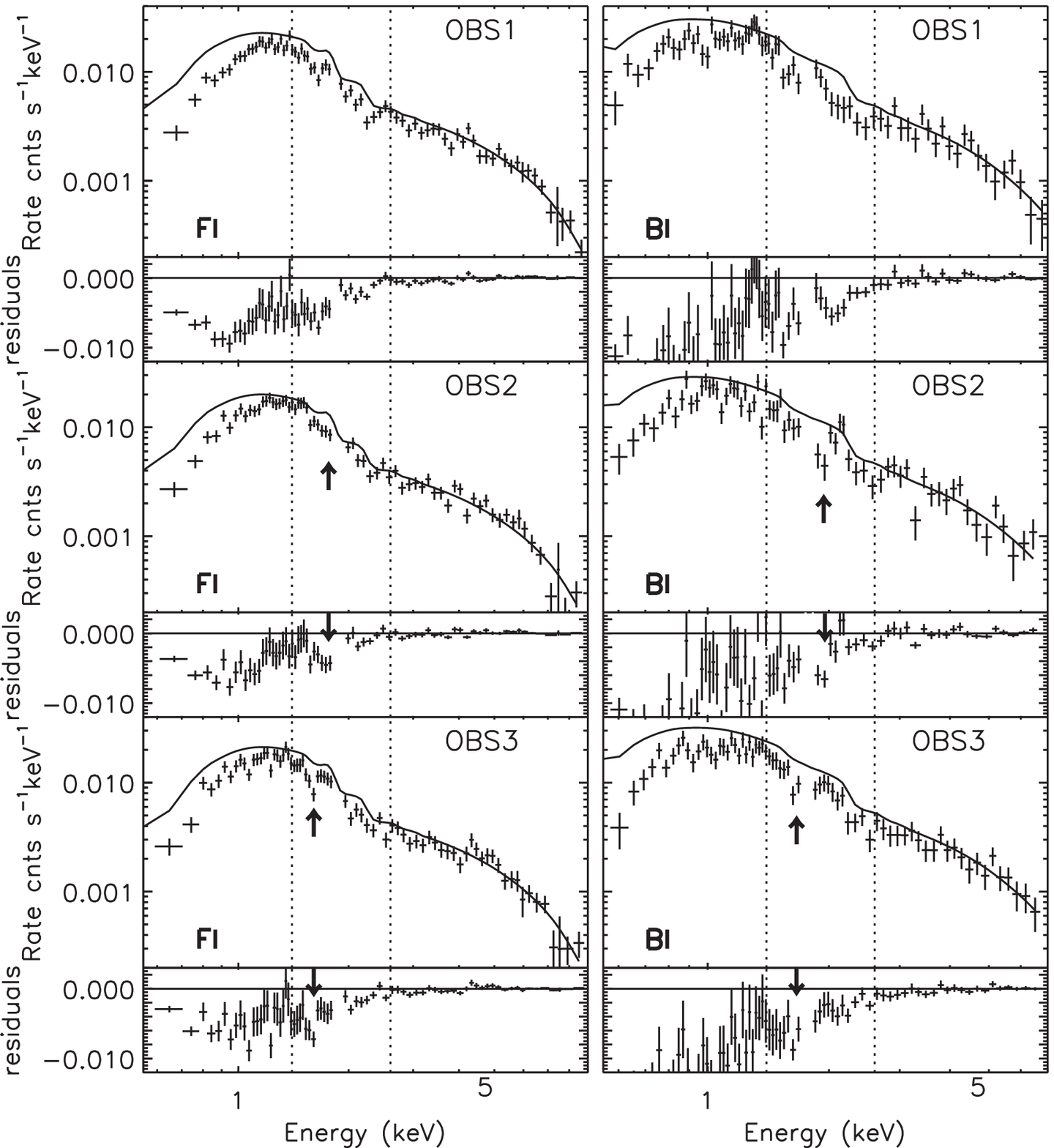}
        \centering
       \caption{ \suzaku\ FI (left panel) and BI (right panel)
        spectra of the combined images of \apm\ for
       the three observations (OBS 1, 2, and 3),
       fit with Galactic absorption and a power-law model to events
       with \OF\ energies above 3.6 keV and then extrapolated to lower energies. In the lower panel of each
       observation, we show the residuals of the fit with 1-$\sigma$ error bars.
      % The best-fit values of the power-law photon indices
       %in these energy ranges for the three observation were within 1-$\sigma$ of $\Gamma=2.0$.
        High-energy absorption features are detected within the \AEn\ range (dotted lines).
        {We have marked with an arrow the best-fitted energies of the
       first absorption feature of model~7 for epochs OBS2 and
       OBS3.}
        For the FI spectra the grouping in OBS1 and OBS2 is 100 counts per bin, and in OBS 3 it is 80 counts per bin.
       For the BI spectra the grouping in OBS1 and OBS2 is 40 counts per bin, and in OBS3 it is 50 counts per bin. }
     \label{fig:FIB}
     \end{figure*}

To illustrate the presence of the high-energy absorption features,
we fit the spectra from observed-frame energies of 3.6--10~keV
with a power-law model and extrapolated this model to the energy
ranges not fit (see Figure~\ref{fig:FIB}). The lower panels in
Figure~\ref{fig:FIB} show the residuals (difference between the
measured counts and model) between the best-fit power-law model
and the FI and BI data, respectively. The best-fit values of the
photon indices in all observations with this model were consistent
with $\Gamma=2$ at the 1-$\sigma$ level. For the purpose of
comparing the absorption residuals between epochs the photon
indices for all observations were set to $\Gamma$ = 2.0. From
these fits we notice that the residuals show an absorption feature
centered near a \RF\ energy of $\sim$8 keV and a possible second
absorption feature near a \RF\ energy of $\sim$10 keV. We fit the
high-energy absorption features with the models listed in
Table~\ref{tab:modn}. From these fits we found that adding to the
APL model an absorption edge (APL+Ed) or two absorption lines
(APL+2AL) improves the fits at the $\gtrsim$99\% confidence level
in the two sets of spectra (FI and BI) and in each observation.
The \Ft\ indicates that we cannot distinguish between the (APL+Ed)
and (APL+2AL) models for fits performed to the spectra of \apm\ in
epochs OBS2 and OBS3, since both models fit equally well the \AEn\
\RF\ absorption during these epochs. \footnote{We only find
marginal improvements in fits to the spectra of \apm\ taken in
epochs OBS2 and OBS3 using model 7 (APL+2AL) compared to fits
using model 6 (APL+Ed). Specifically, in epoch OBS2 these
improvements are at the 61\% and 91\% significance levels in the
FI and BI, and in epoch OBS3 they are at the 68\% and 50\%
significance levels.} However, fits to the FI and BI spectra of
epoch OBS1 using the APL+2AL model provide a significant
improvement at the 95\% and 98\% confidence levels, respectively,
compared to fits that use the APL+Ed model. It is important to
note that the APL+2AL model was clearly favored over the APL+Ed
model in a previous 88.8 ks \chandra\ observation \citep{Cha02}.
Fits to the spectra in epoch OBS1 with a model that includes an
absorption notch (see Table~\ref{tab:modn}) also provide a
significant improvement compared to ones using the APL+Ed model.
These \Ft\ improvements are at the $\sim$98\% and $\sim$99\%
levels of significance in the FI and BI spectra, respectively. We
note, however, that when we compare the quality of the spectral
fits that use the APL+Ed model with fits that use either the
APL+2AL or APL+notch models, the $F$-test may not be a reliable
tool. The reason for the non-reliability of the $F$-test is that
we are not comparing nested models (see Protassov et al. 2002 for
details). In order to check the reliability of the $F$-test for
these cases, we performed Monte Carlo simulations of 10,000 fake
spectra (using the FAKEIT command of XSPEC) assuming an APL+Ed
model. In these simulations, the energy and optical depth of the
absorption edge are assumed to be normally distributed around
their fitted values for epoch OBS1 (see model~6 of
Table~\ref{tab:modn}), with a standard deviation given by the
error bars of the fits. All other parameters of the APL+Ed model
were set to their best-fitted values (epoch OBS1 and model~6 of
Table~\ref{tab:modn}). The results of our Monte Carlo analysis are
presented in Table~\ref{tab:fte}. In each simulation we have
fitted the data with the null model (APL+Ed) and the alternative
model (either APL+2AL or APL+notch). We then calculated the value
of the $F$-statistic between the null model and the alternative
model. In Table~\ref{tab:fte} we show that the $p$-value, which
represents the fraction of simulated cases with values of the
$F$-statistic higher than the actual value obtained from our real
data, is similar to the null probability of the $F$-test. We
therefore conclude that our $F$-test values are reliable and are
approximately representative of the improvement of the alternative
model (either APL+2AL or APL+notch) with respect to the null model
(APL+Ed).

\begin{deluxetable*}
{cccccccccc} \tabletypesize{\scriptsize}
%\tabletypesize{\footnootsize}
%\tabletypesize{\small}
%\tablecolumns{11} \tablewidth{0pt} \tablecaption{$F$-test and
%$p$-test in OBS1 assuming PL+Edge as the null model.
%label{tab:fte}}
\tablecolumns{11} \tablewidth{0pt} \tablecaption{ Estimates of the
improvement of fits to the spectra of APM~08279+5255
using alternative models to the APL+Edge model. \label{tab:fte}}

\tablehead{
 \colhead{Alternative model} & \colhead{Spectrum$^a$} &
\colhead{$F$-statistic/null probability$^b$} & \colhead{$p^c$}}

\startdata

APL+2AL  & FI & 2.59 / 4.4 $\times$ 10$^{-2}$ & 5.2 $\times$ 10$^{-2}$\\
APL+2AL  & BI & 3.05 / 2.3 $\times$ 10$^{-2}$ & 3.1 $\times$ 10$^{-2}$\\
APL+notch  & FI & 5.32 / 2.4 $\times$ 10$^{-2}$ & 6.8 $\times$ 10$^{-2}$ \\
APL+notch  & BI & 7.15 / 0.9 $\times$ 10$^{-2}$ & 3.9 $\times$ 10$^{-2}$ \\

\enddata
\tablenotetext{a}{The XIS FI and BI spectra of APM~08279+5255
considered in this comparison are taken from epoch OBS1.}

\tablenotetext{b}{The value on the left of the slash is the $F$-statistic and is given by
$F=\frac{\chi_{\nu_1}^2-\chi_{\nu_1}^2}{\Delta
\nu}/\frac{\chi_{\nu_2}^2}{\nu_2}$. The value on the right of the slash
represents the probability of exceeding the $F$-statistic based on the $F$-test.}

\tablenotetext{c}{The $p$-value represents the probability of
exceeding the $F$-statistic based on our Monte Carlo simulations.
The value of the $F$-statistic is obtained by comparing the null
model (APL+Edge) with the alternative model listed in the first
column for fits performed to the spectra of APM~08279+5255
obtained in epoch OBS1.}

%is the fraction of trials in a
%Monte-Carlo simulation (10000 trials) where the F-test gets higher
%that the F-test obtained from the data (column 3). The value of
%the $F$-test is obtained comparing the null model (PL+Edge) with
%the alternative model (first column). }

\end{deluxetable*}

\begin{figure}
   \includegraphics[width=8cm]{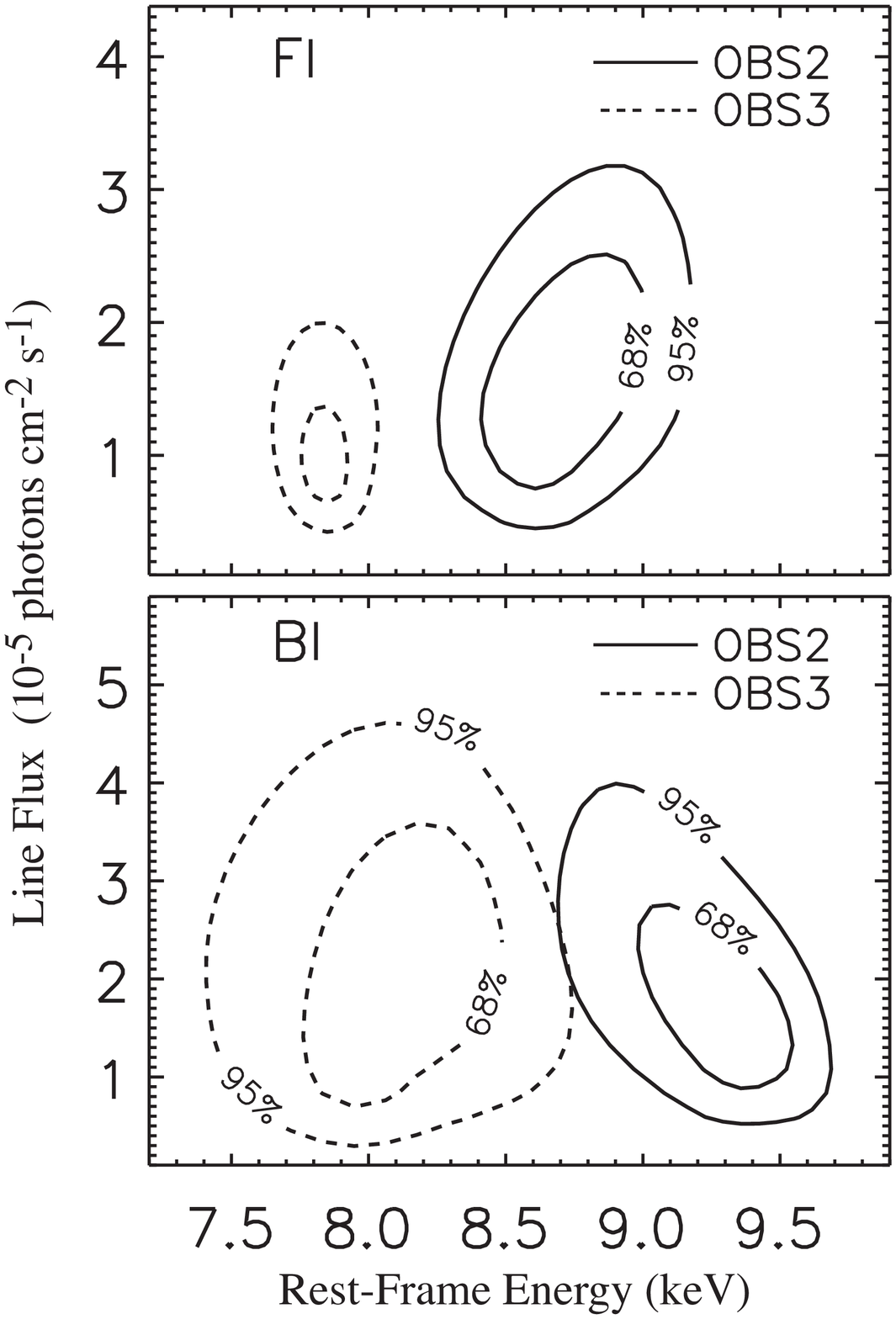}
        \centering
       \caption{68\% and 95\% confidence contours of \hbox{absorption-line} fluxes versus \hbox{absorption-line}
        energies of the first modeled absorption line (model 7, Table~\ref{tab:modn}). The upper and lower
        panel contours correspond to fits performed to the FI and BI spectra, respectively. Solid and
        dotted lines are contours for epochs OBS2 and OBS3, respectively.}
     \label{fig:var}
     \end{figure}

The results of the spectral fits shown in Table~\ref{tab:modn}
indicate a change (greater than 1-$\sigma$ ) of the energies of
the best-fit values of the first absorption line (abs1; model~7),
and in the absorption-edge energy ($E_{\rm Edge}$; model~6)
between epochs OBS2 and OBS3 in both the FI and BI spectra. This
change is also suggested by the residuals in Figure~\ref{fig:FIB},
where we have marked with an arrow the best-fitted energies of the
first absorption feature of model~7 for epochs OBS2 and OBS3. The
shift in the energy of the first absorption line is indicative of
possible variability of the outflow. This change can be seen more
clearly in Figure~\ref{fig:var} where we show the ${\chi}^{2}$
confidence contours of the best-fit energies of the first
absorption line (APL + 2AL model) versus its line-flux
normalization in epoch OBS2 (solid line) and in epoch OBS3 (dotted
line).\footnote{We note that for the FI spectra of \apm\ during
epoch OBS2, the first absorption feature (abs1) falls near the Si
K edge where events with energies lying in the range 1.7--1.95~keV
were ignored in the analysis. Although the loss of these data
points adds a larger statistical error to the best-fitted
parameters of abs1, it does not significantly affect the analysis.
Specifically, this error in the best-fitted energies is less than
the value of $\sigma_{\rm abs1}$ and more likely close to
$\frac{\sigma_{abs1}}{2}\sim0.2~{\rm keV}$ (see
Table~\ref{tab:modn}).} In the upper and lower panels of
Figure~\ref{fig:var}, we show the ${\chi}^{2}$ confidence contours
of the FI and the BI spectra, respectively. The confidence
contours touch at the $\sim$99\% level of significance for the FI
spectra and at the $\sim$95\% level of significance for the BI
spectra. The probabilities that the flux-energy parameters of the
first absorption line (model 7; Table~\ref{tab:modn}) are the same
between OBS2 and OBS3 (null probabilities) are
$\lesssim$$1\times10^{-4}$ and $\lesssim$$2.5\times10^{-3}$ in the
FI spectra and BI spectra, respectively.\footnote{The square of
the probabilities of being outside the confidence contours that
barely touch (see Figure~\ref{fig:var}) is an upper limit to the
null probabilities.} To take into account possible sampling
effects caused by the number of trials used in our variability
analysis we multiply the null probabilities by six. This factor
corresponds to the number of absorption lines (two) times the
number of observations (three). We conclude that the variability
of the first absorption line is significant at the $\gtrsim$99.9\%
and $\gtrsim$98\% levels in the FI and BI spectra, respectively.
\newline \indent We note
that the slight differences ( $ < $ 68\% significance) between the
FI and BI confidence contours may possibly be associated with
differences in the responses, variations in the signal-to-noise
ratios of the two detectors, and statistical noise.

\begin{deluxetable*}
{cccccccccc} \tabletypesize{\scriptsize}
%\tabletypesize{\footnootsize}
%\tabletypesize{\small}
\tablecolumns{11} \tablewidth{0pt} \tablecaption{Results from
spectral fits using XSTAR to epochs OBS1, OBS2 and OBS3 of \apm.
\label{tab:xsta}}

\tablehead{
\colhead{XSPEC Model} & \colhead{Parameter} & &
\multicolumn{2}{c}{OBS1} & \multicolumn{2}{c}{OBS2} &
\multicolumn{2}{c}{OBS3} \\
 & & &  Values FI & Values BI & Values FI & Values BI & Values FI & Values BI
 }

\startdata

XSTAR1........ &  $\Gamma$ & & 1.96$\pm$0.04 & 1.99$\pm$0.05  & 2.00$\pm$0.05 & 1.96$\pm$0.07 & 1.97$\pm$0.05 & 1.95$\pm$0.06\\
&   \lnh & & 23.35$\pm$0.08 & 23.19$\pm$0.10 & 23.30$\pm$0.09 & 23.28$\pm$0.12 & 23.25$\pm$0.09 & 23.26$\pm$0.08 \\
&   \lip & & 1.02$\pm$0.52 & 0.45$\pm$0.36 & 0.99$\pm$0.52 & 1.76$\pm$0.62 & 1.05$\pm$0.75 & 1.51$\pm$0.66\\
&   $\chi^2/\nu$ & & 69.1/71  & 70.7/69  & 82.0/64  & 92.7/64 & 73.8/69 & 48.2/69 \\
&   $P(\chi^2/\nu)^c$ & & 0.54 & 0.42 & 0.06 & 0.01 & 0.32 & 0.97 \\
\\

XSTAR2........ &  $\Gamma$ & & 1.94$\pm$0.05 & 2.00$\pm$0.07 & 2.01$\pm$0.05 & 1.95$\pm$0.07 & 1.97$\pm$0.05 & 1.95$\pm$0.06\\
&   \lnh & & 23.30$\pm$0.09 & 23.18$\pm$0.10 & 23.40$\pm$0.08 & 23.24$\pm$0.12 &  23.25$\pm$0.09 & 23.26$\pm$0.08 \\
&   \lip & & 1.09$\pm$0.51  & 0.54$\pm$0.42 & 1.28$\pm$0.65 & 1.70$\pm$0.78 &  1.05$\pm$0.75 & 1.51$\pm$0.66\\
&   $a_{\rm Fe}$ & & 1.1$\pm$0.3 & 1.8$\pm$0.5 & 0.6$\pm$0.2  & 1.2$\pm$0.5 & 1.0$\pm$0.3 & 1.0$\pm$0.4 \\
&   $\chi^2/\nu$ & & 68.7/70  & 65.2/68 & 77.1/63 & 92.5/63 & 73.8/68 & 48.2/68 \\
&   $P(\chi^2/\nu)^c$ & & 0.52 & 0.57 & 0.11 & 0.01 & 0.29 & 0.97 \\
\\

%XSTAR2........ & $\Gamma$ & & $1.95\pm0.05$ & $1.95\pm0.07$ \\
%&   \lnh & & $23.33\pm0.08$ &  $23.21 \pm 0.10$ \\
%&   \lip & & $<$0.22  &  $<$0.57  \\
%&  $E_{\rm line}$[keV] & & $10.85\pm0.21$ & $10.62\pm0.37$ \\
%&  $\sigma_{line}$[keV] & & $<$0.55 &  $0.98\pm0.78$\\
%&   $\chi^2/\nu$  &  &  61.2/68  & 57.6/66   \\
%\\

XSTAR3........ & $\Gamma$ & &   1.97$\pm$0.05 & 1.96$\pm$0.08 &  2.02$\pm$0.05  & 1.95$\pm$0.07 & 1.96$\pm$0.05 & 1.92$\pm$0.06    \\
&   \lnh$_{\rm abs1}$ & & 23.31$\pm$0.09 & 23.17$\pm$0.18 & 23.40$\pm$0.08  & 23.28$\pm$0.10 & 23.21$\pm$0.09 & 23.14$\pm$0.09 \\
&   \lip$_{\rm abs1}$ & & 1.17$\pm$0.55  & 0.50$\pm$0.36  & 1.26$\pm$0.47 & 1.28$\pm$0.25 & 0.78$\pm$0.46 & 1.31$\pm$0.49  \\
&    $a_{\rm Fe}$ & & 0.8$\pm$0.2 & 1.5$\pm$0.7 & 0.6$\pm$0.3 & 1.2$\pm$0.6 & 1.1$\pm$0.3 & 1.3$\pm$0.5\\
&   z$_{\rm abs1}$  & & 3.91 & 3.91 & 3.91 & 3.91 & 3.91 & 3.91 \\
&   \lnh$_{\rm abs2}$ & & 23.07$\pm$0.31 & 23.36$\pm$0.26 & 22.79$\pm$0.36 & 22.87$\pm$0.41 & 22.91$\pm$0.35 & 23.21$\pm$0.39  \\
&   \lip$_{\rm abs2}$ & & 3.7$\pm$0.4  & 3.6$\pm$0.5 & 3.5$\pm$0.3  & 3.8$\pm$0.5  & 3.5$\pm$0.4  & 3.7$\pm$0.5   \\
&   z$_{\rm abs2}$ & & 2.06$\pm$0.05 & 2.18$\pm$0.08 & 1.98$\pm$0.07& 1.79$\pm$0.12 & 2.19$\pm$0.13 & 2.07$\pm$0.10\\
&   $\chi^2/\nu$  & & 58.0/67  & 54.9/65 & 73.6/60 & 89.8/60 & 70.2/65  & 43.6/65 \\
&   $P(\chi^2/\nu)^c$ & & 0.77 & 0.81 & 0.11 & $8\times10^{-3}$ & 0.31 & 0.98\\
\\

%XSTAR3a........ & $\Gamma$ & &   1.96$\pm$0.05 & 1.96$\pm$0.08 \\
%&   \lnh$_{\rm abs1}$ & & 23.24$\pm$0.09 & 23.19$\pm$0.26 \\
%&   \lip$_{\rm abs1}$ & & 0.98$\pm$0.55  & 0.48$\pm$0.36  \\
%&   z$_{\rm abs1}$  & & 3.91 & 3.91\\
%&   \lnh$_{\rm abs2}$ & & 22.90$\pm$0.16 & 23.29$\pm$0.26 \\
%&   \lip$_{\rm abs2}$ & & 3.5  & 3.5  \\
%&   z$_{\rm abs2}$ & & 2.06$\pm$0.05 & 2.18$\pm$0.08\\
%&   $\chi^2/\nu$  & & 59.2/69  & 57.1/67   \\
%\\

XSTAR4........ & $\Gamma$ & &  1.94$\pm$0.06 & 1.95$\pm$0.07 & 2.01$\pm$0.05 & 1.92$\pm$0.07 & 1.93$\pm$0.05 & 1.91$\pm$0.07 \\
&   \lnh$_{\rm abs1}$ & & 22.86$\pm$0.16 & 23.01$\pm$0.26 & 22.93$\pm$0.31 & 23.16$\pm$0.36 & 22.82$\pm$0.21 & 23.02$\pm$0.28 \\
&   \lip$_{\rm abs1}$ & & 3.8$\pm$0.3  & 3.5$\pm$0.2 & 3.6$\pm$0.2  & 3.8$\pm$0.5 & 3.1$\pm$0.3  & 3.3$\pm$0.3 \\
&   z$_{\rm abs1}$  & & 3.08$\pm$0.10 & 3.22$\pm$0.12 & 2.78$\pm$0.08 & 2.57$\pm$0.12 & 3.24$\pm$0.06 &  3.12$\pm$0.08 \\
&   ($\frac{v_{\rm abs1}}{c}$)$^b$ & & (0.19$\pm$0.02) & (0.16$\pm$0.03) & (0.27$\pm$0.02) & (0.32$\pm$0.03) & (0.15$\pm$0.01) & (0.18$\pm$0.02)   \\
&   \lnh$_{\rm abs2}$ & & 23.38$\pm$0.15 & 23.43$\pm$0.22 & 22.91$\pm$0.35 & 23.09$\pm$0.39 & 23.10$\pm$0.38 & 23.07$\pm$0.41\\
&   \lip$_{\rm abs2}$ & & 3.6$\pm$0.5  & 3.4$\pm$0.2 & 3.4$\pm$0.3  & 3.7$\pm$0.6 & 3.5$\pm$0.3  & 3.6$\pm$0.6 \\
&   z$_{\rm abs2}$ & & 2.10$\pm$0.06 & 2.17$\pm$0.08 & 1.97$\pm$0.09 & 1.78$\pm$0.13 & 2.18$\pm$0.12 & 2.05$\pm$0.10  \\
&   ($\frac{v_{\rm abs1}}{c}$)$^b$  & & (0.45$\pm$0.02) & (0.43$\pm$0.02) & (0.48$\pm$0.02) & (0.53$\pm$0.03) & (0.42$\pm$0.03) & (0.46$\pm$0.03)  \\
&   $\chi^2/\nu$  & & 56.3/66  & 55.1/64  & 65.1/59 & 80.0/59 & 67.1/64 & 43.4/64 \\
&   $P(\chi^2/\nu)^c$ & & 0.80 & 0.78 & 0.27 & 0.04 & 0.37 & 0.98\\
\\

\enddata

\tablenotetext{a}{ XSTAR1$\equiv$ XSPEC model warmabs(pow);
XSTAR2$\equiv$ XSPEC model warmabs(pow) (FeA variable);
XSTAR3$\equiv$ XSPEC model warmabs*warmabs(pow); XSTAR4$\equiv$
XSPEC model wabs*zwabs*warmabs*warmabs(pow) }

\tablenotetext{b}{$\frac{v_{\rm abs1}}{c}$ and $\frac{v_{\rm
abs2}}{c}$ corresponds to the estimated outflow velocities. They
are calculated from equation \ref{eq:Dop} based on the redshift of
the absorbers of model XSTAR4. They are not parameters of the
spectral fit.}

\tablenotetext{c}{Probability that $\chi^2/\nu$ is greater than
the value obtained.}

\end{deluxetable*}

\subsubsection{XSTAR spectral fits.} \label{XSTAR}

The spectral analysis presented in \S\ref{XSPEC} indicates that
the intrinsic X-ray absorbing medium of \apm\ is complex and
contains absorbers with different properties (see models 5--7 in
Table~\ref{tab:modn}). We identified a \LI\ absorber with a column
density of \lnh$\sim$23 and an ionization parameter of
\lip$\lesssim$0~\footnote{The spectral fits did not show an
improvement using a warm-absorber model; however the
$\lesssim$4~keV absorption was not well constrained since the
\suzaku\ spectra start at \RF\ energies of $E \sim 2$~keV.} (see
models 2--4 in Table~\ref{tab:modn}). This \LI\ absorber is
required to model the absorption detected below $\sim$2~keV
(observed-frame). An additional complex absorber is required to
fit the broad absorption features with rest-frame energies between
\AEn. Given the range of energies and variability of the broad
absorption features, we interpret this absorption as a blend of
highly ionized (2.75$\lesssim$\lip$\lesssim$4)  iron absorption
lines blueshifted by an outflow. This explanation is consistent
with recent models that attempt to simulate X-ray BALs in quasars
\citep[e.g.,][]{Sch07}. To test this interpretation we next employ
more complex, but more realistic, models to fit the \apm\ spectra.

As a first attempt we fit the low and high-energy absorption of
\apm\ in epoch OBS1 with a model that includes a power-law (with
Galactic absorption) and one warm absorber (model XSTAR1,
Table~\ref{tab:xsta}). The warm-absorber model is calculated using
the XSTAR code \citep[see, e.g.,][]{Kal01, Kal96}. XSTAR
calculates the physical conditions and absorption-emission spectra
of photoionized gases with variable abundances. In the current
analysis we use a recent implementation of the XSTAR code called
WARMABS that can be used as a model within XSPEC. For the WARMABS
model we assume turbulent velocities $v_{\rm turb}=1,000~\kms$
(default velocity of the model).\footnote{Since a warm absorber
with $\lip \sim 1.0$ is expected to have a temperature of
$\lesssim10^6$~K \cite[e.g.,][]{Che03}, we do not expect a thermal
broadening higher than 100~\kms.} For the fits we assumed solar
abundances, a redshift of 3.91 for the warm absorber, and we left
the column density and ionization parameter of the warm absorber
free to vary in the fit. We note that spectral fits using model
XSTAR1 attempt to fit both the the low and high-energy absorption
of \apm\ with a single warm absorber.
\newline \indent To constrain the iron abundance ($a_{\rm Fe}$), we allowed
this parameter to vary in the spectral fits in the model XSTAR2.
The only difference between models XSTAR1 and XSTAR2 is that
$a_{\rm Fe}$ is fixed in model XSTAR1 and free to vary in model
XSTAR2. We find that allowing $a_{\rm Fe}$ to vary in our spectral
fits does not lead to a significant improvement in the fits, and
the best-fit values of $a_{\rm Fe}$ in fits with model XSTAR2 are
consistent with no iron over-abundance. Our results do not confirm
an apparent iron over-abundance ($a_{\rm Fe}$$>$2) claimed by
\cite{Has02} and \cite{Ram08} based on their analyses of previous
observations of \apm.

We next assumed the spectral model XSTAR3 consisting of a
power-law,  one stationary ionized absorber with a turbulent
velocity of $v_{\rm turb}=1,000~\kms$ and a second  outflowing
ionized absorber with a turbulent velocity of $v_{\rm
turb}=10,000~\kms$ (the maximum value  allowed by the
model\footnote{The high-enengy absorption features modeled with
Gaussian absorption lines in model 7 could be the result of one or
more highly ionized absorbers. The Doppler broadening velocities
of each absorption line component  of model 7 are
$\sim$$\frac{\sigma_{\rm absi}}{E_{\rm absi}}$ (where $i$=1,2
indicates the component). From Table~\ref{tab:modn} these Doppler
broadening velocities are at first order comparable to the assumed
values of $v_{\rm turb}$.}). We allowed the ionization parameters
of both ionized absorbers and the redshift of the second ionized
absorber to vary in the fit (model XSTAR3; Table~\ref{tab:xsta}).
We find that the best fitted  redshift of the second warm absorber
is $z\sim2$ in both the FI and BI spectra. The \Ft\ indicates an
improvement in the fits of OBS1 with  model XSTAR3, that assumes
two ionized absorbers, compared to fits with models XSTAR1 and
XSTAR2, that assume a single ionized absorber, at the
$\gtrsim$99.5\% of significance level  in the FI and BI spectra.
We conclude that a single warm-absorber model cannot accurately
fit both the low and high-energy absorption in \apm.

We finally assumed the spectral model XSTAR4 consisting of an
absorbed power-law and two outflowing ionized absorbers. We
assumed turbulent velocities of $v_{\rm turb}=10,000~\kms$ for the
first and second outflowing ionized absorber (model XSTAR4;
Table~\ref{tab:xsta}). The main difference between models XSTAR3
and XSTAR4 is that the redshift of the first warm-absorber is
fixed in model XSTAR3 to the  systemic redshift of the quasar and
free to vary in model XSTAR4 and that model XSTAR4 includes a
neutral absorber. For fits using model XSTAR4, we allow the
redshifts, column densities, and ionization parameters of the
absorbers to vary. The $\chi^2$ values for these fits are similar
to those found for model 7 (see Table~\ref{tab:xsta} and
\ref{tab:modn}). We find on average best-fit redshifts of $z_{\rm
abs1}\sim3$ ($v_{\rm abs1}\sim 0.2c$) and $z_{\rm abs2}\sim2$
($v_{\rm abs1}\sim 0.5c$), and column densities of $\log$ $N_{\rm
H,abs1}\sim23.0$ and $\log$ $N_{\rm H,abs2}\sim23.2$, where $abs1$
and $abs2$ correspond to the two warm absorbers. We confirm the
results of model 7, by finding a significant change in the
best-fitted redshift of the first warm-absorber component in
model~XSTAR4 between epochs OBS2 and OBS3 (see
Table~\ref{tab:xsta}). Even though the two warm-absorber model
results in acceptable fits, the best-fit parameters should only be
considered as basic estimates of the wind properties since the
kinematic and ionization structure of the outflow are likely to be
more complicated. Our spectral fitting results of models that
include ionized absorbers (see Table~\ref{tab:xsta}) indicate that
both models XSTAR3 and XSTAR4 provide acceptable fits to the
spectra of \apm\ for epochs OBS1 and OBS3, however, model XSTAR4
provides a better fit to the data for epoch OBS2 than model
XSTAR3. The \Ft\ indicates that in epoch OBS2, spectral fits using
model XSTAR4 provide an improvement over fits using model XSTAR3
at the $\sim99\%$ level of significance in the FI and BI spectra.

\subsection{PIN spectral analysis.}

We also examined the spectrum of \apm\ in the 10--40 keV energy
band using the PIN-HXD data. Unfortunately, no signal from the
source above the background in the NXB was found. The
background-subtracted source spectrum \footnote{The
background-subtracted source spectrum is obtained by subtracting
the non-X-ray-background and an estimate of the X-ray background
from the total detected PIN spectrum} was found to be within 5\%
of the non-X-ray background (NXB) spectrum provided by the
\suzaku\ team \citep[e.g.,][]{Miz07}. We arrive at a similar
conclusion from the count rates presented in Table~\ref{tab:hxd}.
The non-detection with the PIN provides an upper limit on the flux
density at 20~keV of \apm\ of $\sim10^{-3}$photons s$^{-1}~{\rm
keV}^{-1}$. This limit is consistent with an extrapolation of the
XIS spectrum.

\section{DISCUSSION}

The \Rx\ spectrum of \apm\ is known to contain absorption features
at rest-frame energies above 7~keV \citep{Cha02, Has02}. These
features have been interpreted in the past in two different ways.
The first interpretation by \cite{Cha02} was based primarily on
the analysis of the 2002 \chandra\ observation of \apm\ and posits
that the absorption features are due to highly blueshifted Fe XXV
K$\alpha$ and/or Fe XXVI K$\alpha$ absorption lines. An {\sl
XMM-Newton} observation of \apm\ performed 1.8 weeks (proper-time)
after the {\sl Chandra} observation showed significantly different
high-energy absorption structure which was interpreted by
\cite{Cha03} to imply variability of the absorption features over
timescales of the order of weeks. The second interpretation by
\cite{Has02} is based on the 2002 {\sl XMM-Newton} observation of
\apm\ and proposes that the high-energy absorption feature arises
from an iron absorption edge produced by a metal enriched (Fe/O
$\approx$ 2--5 Fe/O$|_\odot$) ionized absorber. One important
conclusion from section \ref{XSTAR} is that the absorption feature
found at \AEn\ \RF\ can be fitted with two highly ionized
blue-shifted warm absorbers that do not require super-solar
metallicities. We note that \cite{Has02} and \cite{Ram08} had
claimed iron over-abundance ($a_{\rm Fe}$$>$2) based on their
analyses of previous observations of \apm. In support of the
two-component ``iron-blend-outflow'' scenario we mention that the
two absorption-line model and the notch model (models 7 and 5;
Table~\ref{tab:modn}) provide significantly better fits to the
absorption feature between \AEn\ \RF~than an absorption-edge model
in epoch OBS1 (see \S 3). We also note that, in the 2002 $\sim$90
ks \chandra\ observation analyzed in \cite{Cha02}, the model
containing two absorption lines successfully fits the \AEn\ \RF\
feature whereas an absorption-edge model did not provide an
acceptable fit. The absorption described either by two absorption
lines or a notch, may crudely represent absorption through an
outflow with a large velocity gradient along the flow. Variability
of the kinematic and ionization state of the outflow may explain
why, depending on the observation, these absorption features could
be modelled by either a notch, absorption lines, or an edge
\citep[see figure 4 of][]{Sch07}. We conclude that a time-variable
outflow provides a plausible explanation for all the past \Rx\
observations of the absorption features of \apm\
\citep{Cha02,Has02} including those analyzed here. In
\S\ref{s:var} we provide plausible explanations for the observed
variability of the high-energy absorption. In \S\ref{s:kin} we use
the results of our spectral analysis to place constraints on the
kinematics of the outflow and in \S\ref{s:out} we provide
estimates of the mass-outflow rate and efficiency of the outflow
of \apm.

\begin{figure*}
   \includegraphics[width=16cm]{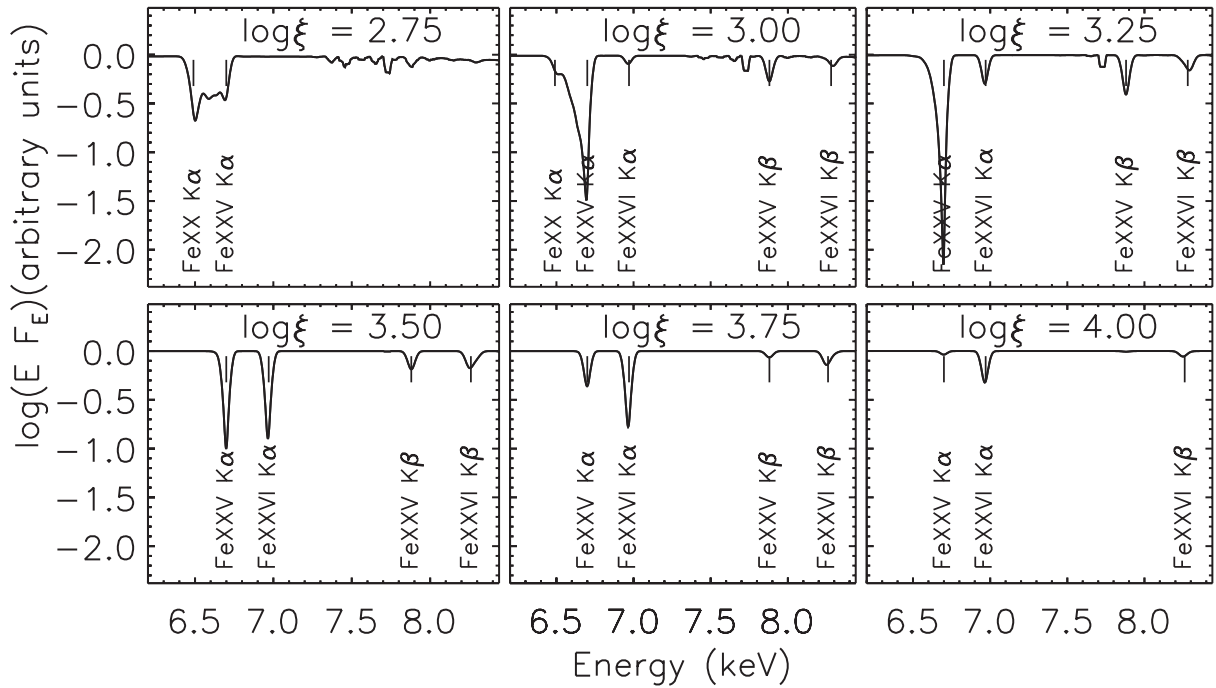}
        \centering
      \caption{Simulated 6--8~keV absorbed spectra (${\rm log}(E F_{\rm E}$))
      assuming an incident power-law spectrum with a photon index of $\Gamma$=2.
      %produced from a model consisting of a power-law ($\Gamma$=2) with an ionized absorber.
      The ionized absorber is modeled with the XSTAR model WARMABS
      assuming a column density of
       $\lnh=23$, solar abundances, $v_{\rm turb}=1000~\kms$, and the following six
       values of the ionization parameter
       \lip=2.75, 3.00, 3.25, 3.50, 3.75, and 4.0, respectively. Some of the main absorption
       lines in this range of ionization states have been marked in the figure. Among them
       \FeXX\ K$\alpha$ ($2s^22p^3-1s2s^22p^4$; 6.50~keV), \FeXXV\ K$\alpha$ ($1s^2-1s2p$; 6.70~keV),
       \FeXXV\ K$\beta$ ($1s^2-1s3p$; 7.89~keV), \FeXXVI\ K$\alpha$ ($1s-2p$; 6.97~keV),
       \FeXXVI\ K$\beta$ ($1s-3p$; 8.27~keV).}
     \label{fig:lin}
     \end{figure*}

\subsection{Origin of the Variability of the High-Energy Absorption
Feature.} \label{s:var}

Assuming the first interpretation of the origin of the high-energy
absorption features in \apm\ the observed shift in the energy of
the first absorption line between epochs OBS2 and OBS3 is likely
due to a change in the outflow velocity of the absorber. Two
alternative explanations of the shift are a change in the
direction of the outflow (with respect to the line of sight) and a
variation in the ionization parameter of the absorber.  A change
in the direction of the outflow is expected to show a shift in
energy of both components of model~7 (Table~\ref{tab:modn}). We
only find variability in one component (abs1, see \S3); however
this picture could still be valid if each outflow component is
driven independently. A change in the ionization parameter seems
to be a less probable scenario. We checked this by fitting the
spectra of epochs OBS2 and OBS3 simultaneously with model XSTAR4,
keeping the redshift of abs1 as the only common parameter between
the fits. In the case where a change in the ionization parameter
produced the detected variability of the first absorption line
abs1, we expect that the fits to the spectra of epochs OBS2 and
OBS3 will not be improved by allowing the redshift of abs1 to vary
independently in these fits. For the simultaneous fit to the
spectra of epochs OBS2 and OBS3, where the redshift of abs1 was
kept a common parameter, we find $z_{\rm abs1}{\rm
(FI)}=3.11\pm0.04$, $z_{\rm abs1}{\rm (BI)}=3.04\pm$0.06,
$\chi^2_{\rm FI}/\nu=138.1/124$ and $\chi^2_{\rm
BI}/\nu=136.3/124$. In the case where we fit the spectra of epochs
OBS2 and OBS3 independently using model XSTAR4 (see
Table~\ref{tab:xsta}) we find $\chi^2_{\rm FI}/\nu=132.2/123$ and
$\chi^2_{\rm BI}/\nu=123.4/123$.\footnote{These values are
obtained by summing the $\chi^2$ and degrees of freedom of epochs
OBS2 and OBS3 in Table~\ref{tab:xsta}.} The improvement based on
the \Ft\ of fitting the spectra of epochs OBS2 and OBS3
independently, compared to keeping a common redshift of abs1, is
at the $\sim$99\% and $>$99.9\% level of significance in the FI
and BI spectra. We conclude that the variability of the energy of
the first absorption line abs1 is likely not driven by changes in
the ionization parameter of abs1. {\sl Suzaku} cannot resolve the
images of \apm, however, the time-delays between the two brightest
images A and B of \apm\  is estimated to be of the order of a few
hours \citep[e.g.,][]{Mun01}, much shorter than the observed
variability of the high-energy absorption feature. We therefore do
not expect the combined X-ray spectrum of all images of \apm\ to
differ from that of the individual images within the time-delay.

\begin{deluxetable*}
{cccccccccccc} \tabletypesize{\scriptsize}
%\tabletypesize{\footnootsize}
%\tabletypesize{\small}
\tablecolumns{12} \tablewidth{0pt} \tablecaption{
The minimum and maximum
energies and velocities of the high-energy absorption features in \apm.
\label{tab:vma}}

\tablehead{
\colhead{Model$^a$} & \colhead{OBS} &
\colhead{Instrument} & & \colhead{$E_{\rm min}$} &
\colhead{$E_{\rm max}$} &
\colhead{\vmin} & \colhead{\vmax} \\
& & & & [keV] & [keV] & [$c$] & [$c$]
}

\startdata

5....... & 1 & XIS FI & & 7.60$\pm$0.24 & 11.58$\pm$0.24 & 0.13$\pm$0.03 & 0.52$\pm$0.02 \\

7....... & 1  & XIS FI & &7.52$\pm$0.36 & 12.16$\pm$0.52 & 0.12$\pm$0.05 & 0.56$\pm$0.04 \\

5....... & 2 & XIS FI & & 7.82$\pm$0.37 & 11.68$\pm$0.37 & 0.16$\pm$0.05 & 0.52$\pm$0.03 \\

7....... & 2 & XIS FI & & 7.88$\pm$0.48 & 11.57$\pm$0.51 & 0.17$\pm$0.06 & 0.51$\pm$0.04 \\

5....... & 3 & XIS FI & & 7.38$\pm$0.34 & 11.42$\pm$0.34 & 0.10$\pm$0.05 & 0.50$\pm$0.03 \\

7....... & 3 & XIS FI & & 7.51$\pm$0.32 & 11.70$\pm$1.06 & 0.12$\pm$0.05 & 0.52$\pm$0.08 \\

5....... & 1 & XIS BI & &  7.46$\pm$0.53 & 12.17$\pm$0.53 & 0.11$\pm$0.08 & 0.56$\pm$0.04\\

7....... & 1  & XIS BI & &  7.24$\pm$0.51 & 12.47$\pm$1.12 & 0.08$\pm$0.07 & 0.58$\pm$0.08\\

5....... & 2 & XIS BI  & & 7.91$\pm$0.78 & 12.92$\pm$0.78 & 0.18$\pm$0.10 & 0.61$\pm$0.05\\

7....... & 2 & XIS BI  & & 8.74$\pm$0.58 & 12.24$\pm$1.16 & 0.28$\pm$0.07 & 0.56$\pm$0.08\\

5....... & 3 & XIS BI   & & 7.70$\pm$0.63 & 12.22$\pm$0.63 & 0.15$\pm$0.09 & 0.56$\pm$0.04\\

7....... & 3 & XIS BI   & & 7.53$\pm$0.82 & 12.40$\pm$0.92 & 0.12$\pm$0.11 & 0.57$\pm$0.06\\

7....... & Cha02$^{b}$ & ACIS BI& & 7.95$\pm$0.11 & 10.28$\pm$0.22 & 0.18$\pm$0.01 & 0.40$\pm$0.02  \\

7....... & Has02$^{b}$ & EPIC pn& & 6.95$\pm$0.44 & 15.00$\pm$1.06 &  $<$0.10 & 0.72$\pm$0.05\\

\enddata

\tablenotetext{a}{Model used to estimate $E_{\rm min}$ and $E_{\rm
max}$. Model 5 (see Table~\ref{tab:modn}) is a power-law with
Galactic absorption, intrinsic absorption, and a notch absorber;
Model 7 (see Table~\ref{tab:modn}) is a power-law with Galactic
absorption, intrinsic absorption, and two absorption lines.}

\tablenotetext{b}{In this Table we identify as Cha02 the 88.8 ks
observation of \apm\ performed with \chandra\ in 2002 and analyzed
in detail in \cite{Cha02}. We also identify as Has02 the 100.2 ks
observation of \apm\ performed with \xmm\ in 2002 and analyzed in
detail in \cite{Has02}.}

\end{deluxetable*}

\subsection{Constraints on the Kinematics of the Outflow.}  \label{s:kin}

Under the premise of the outflow interpretation to explain the
absorption at \RF\ \AEn\ we expect a continuous distribution of
outflow velocities. This range of velocities leads to the the
Doppler shift of the energies of the resonance absorption lines.
The absorption-line rest-frame energies $E_{\rm lab}$ will thus be
shifted to the observed energies $E_{\rm obs}$ according to

\begin{equation}  \label{eq:Dop}
E_{lab}/E_{obs} =\gamma(1-\beta cos\theta),
\end{equation}
where $\gamma$ is the Lorentz factor, $\theta$ is the angle
between the wind and our line of sight (l.o.s), and $\beta=v/c$.
\newline \indent The minimum and maximum projected velocities ($v_{\rm
min}$, $v_{\rm max}$) of the outflow are estimated from the
minimum and maximum energy ranges ($E_{\rm min},E_{\rm max}$) of
the high-energy absorption features in \apm. We obtained $E_{\rm
min}$ and $E_{\rm max}$ from our spectral fits assuming first the
two absorption-line (APL+2AL) model and second assuming the notch
(APL+No) model. Specifically, based on the best-fit values of the
APL+2AL model (model 7; Table~\ref{tab:modn}), we obtain $E_{\rm
min}=E_{\rm abs1}-2\sigma_{\rm abs1}$ and $E_{\rm max}=E_{\rm
abs2}+2\sigma_{\rm abs2}$. From the best-fit values of the APL+No
model (model 5; Table~\ref{tab:modn}) we have $E_{\rm min}=E_{\rm
notch}-W_{\rm notch}/2$ and  $E_{\rm max}=E_{\rm notch}+W_{\rm
notch}/2$. The values of $E_{\rm min}$ and $E_{\rm max}$ are
presented in Table~\ref{tab:vma}. In this table the values of
$E_{\rm min}$ and $E_{\rm max}$ are shown separately for the FI
and BI spectra and for the two different models used to obtain
them (model 7$\equiv$APL+2AL; model 5 $\equiv$APL+No).

\begin{figure*}
   \includegraphics[width=16cm]{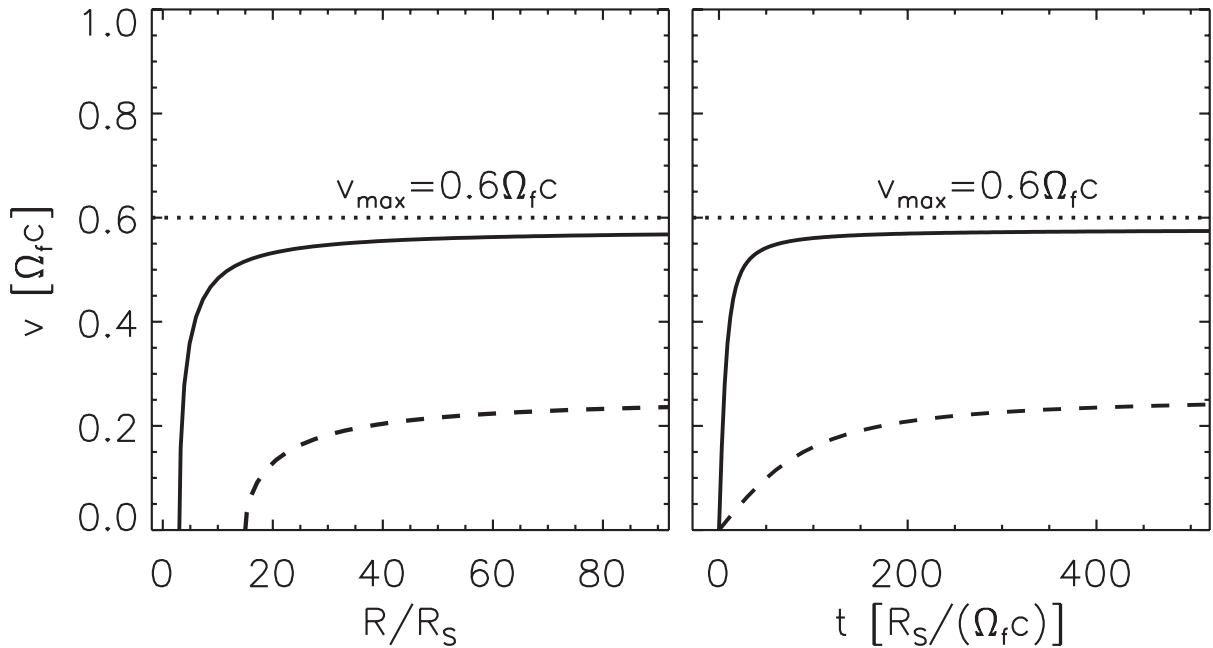}
        \centering
       \caption{Wind velocity (in units of $\Omega_fc$) plotted as a function of radius from the central source (left panel) and as a function of time (right panel)
        for a \hbox{radiation-pressure} driven wind. Notice that the radius is in units of the
        Schwarzschild radius ($R_S=\frac{2GM}{c^2}$)
        and the time is in units of $\tau=\frac{R_S}{\Omega_fc}$.
        The wind velocities are calculated for $R_{\rm launch}=3R_S$ and $R_{\rm launch}=15R_S$, respectively.}
    \label{fig:vrc}
     \end{figure*}

The relative strengths of the iron resonance absorption lines will
depend on the ionization state of the outflowing medium. To
demonstrate this effect in a basic way we have performed several
simulations using the warm-absorber model (WARMABS) of XSTAR. In
Figure~\ref{fig:lin} we show a simulated absorbed spectrum in the
6--8 keV \RF~energy range for an absorbing medium with solar
composition, \lnh=23, and having six different values of the
ionization parameter (\lipR). The two strongest iron lines for
this highly ionized absorbing medium have rest (or laboratory)
energies of 6.70 keV (\FeXXV\ K$\alpha$; $1s^2-1s2p$) and 6.97 keV
(\FeXXVI\ K$\alpha$ $1s-2p$). In general the \FeXXV\ K$\alpha$
line will be stronger than the \FeXXVI\ K$\alpha$ line for a
medium with \lipR. Therefore the absorption at the lower end of
the \AEn\ \RF\ range is most likely associated with the \FeXXV\
K$\alpha$ line. Based on our interpretation of the high-energy
absorption features we estimate $v_{\rm min}$ assuming that the
absorption at the low end of the X-ray BAL is due to a line
arising from highly blueshifted \FeXXV\ K$\alpha$ ($E_{\rm
lab}=6.7$~keV). On the other hand to estimate $v_{\rm max}$ we
make the conservative assumption that the absorption at the high
end of the X-ray BAL is due to a line arising from highly
blueshifted \FeXXVI\ K$\alpha$ ($E_{\rm lab}=6.97$~keV). The
estimated values of $v_{\rm min}$ and $v_{\rm max}$ obtained
through the procedure outlined above are presented in
Table~\ref{tab:vma}. These velocities are obtained using equation
(\ref{eq:Dop}) assuming that our line-of-sight makes an angle of
$\theta=20^{\rm o}$ with the velocity of the outflow
\citep{Cha07}. A change of 10$^\circ$ in $\theta$ will introduce a
variation of $\lesssim$10\% in our estimates of the velocities.
The velocities of the outflow range between 0.1$c$ and 0.6$c$. The
mean value of $E_{\rm max}$ from all CCDs and observations is
12.04$\pm$0.22 keV corresponding to a mean value of $v_{\rm max}$
= 0.55$\pm$0.02$c$. This maximum value of the outflow velocity
constrains the angle between the outflow direction and our line of
sight to be $ < $ 36$^{\rm o}$.\footnote{The Doppler-shift formula
(equation \ref{eq:Dop}) predicts that given a fixed ratio of
$E_{\rm lab}$/$E_{\rm obs}$$\equiv$$R_{\rm lo}$ the maximum angle
between our line of sight and the wind direction is given by
$\theta_{\rm max}={\rm cos}^{-1}(\sqrt{1-R_{lo}^2})$.}  This
relatively small angle is consistent with outflow models that
posit that BAL quasars are viewed through collimated outflows.

As argued in \cite{Cha02, Cha03, Cha07} we are likely observing
the X-ray absorbers as they are accelerated near their launching
radii. We use the following equation to describe the basic
dynamics of a radiation-driven outflow \citep[e.g., equation 1
of][]{Cha02}:

\begin{equation} \label{eq:vwi}
v_{\rm wind}=\left[\right( \Gamma_{\rm f} \frac{L_{\rm UV}}{L_{\rm Edd}}-1
\left) \right ( \frac{1}{R_{\rm launch}}-\frac{1}{R} \left ) \right]^{1/2}
\end{equation}
where $v_{\rm wind}$ is the outflow velocity in units of $c$,
$\Gamma_f$ is the force multiplier, $L_{\rm Edd}$ is the Eddington
luminosity, $R_{\rm launch}$ is the radius (units of $R_S$) at
which the wind is launched from the disk, and $R$ is the distance
(units of $R_S$) from the central source. The expression for the
dynamics of the outflow can be simplified by defining
$\Omega_f=\sqrt{\Gamma_{\rm f} \frac{L_{\rm UV}}{L_{\rm Edd}}-1}$.
%and $R_{\rm launch}$.

In Figure~\ref{fig:vrc} we plot wind velocity versus radius (left
panel) and wind velocity versus time (right panel) for an outflow
launched at radii of $3 R_S$ (solid lines) and $15 R_S$ (dashed
lines). In Figure~\ref{fig:vrc} the units of velocity, distance,
and time are $\Omega_f c$, $R_S$, and $R_S/(\Omega_f c)$,
respectively. {Assuming $L_{\rm UV}\sim 0.2L_{\rm bol}$
\citep{Irw98}, $L_{\rm bol}/L_{\rm Edd}\sim0.2$ and
$\Gamma_f\sim100$ \citep[e.g.,][]{Ara94,Lao02}, we have
$\Omega_f\sim1.7$. $L_{\rm bol}/L_{\rm Edd}\sim0.2$  is obtained
using the $L_{\rm bol}/L_{\rm Edd}$ vs. $\Gamma$ correlation found
in RQ quasars \citep[e.g.,][]{Wan04,She06,She08} for
$\Gamma\sim1.9$. Therefore with $L_{\rm
bol}=7\times10^{15}\mu_L^{-1}L_\odot$
\citep{Irw98,Rei08}\footnote{We note that an estimation of $L_{\rm
bol}$ based on the optical and UV spectra should be more precise
than an extrapolation of $L_{\rm bol}$ based on \Rx\ luminosities
as it is done in \cite{Ram08}.}  $M_{\rm BH}\sim 10^{12}M_\odot
\mu_L^{-1}$, where $\mu_L\sim100$ \citep{Ega00}\footnote{See,
however, \cite{Rei08} that find a magnification of $\mu_L\sim4$.
\cite{Rei08} also use the observed width of the CIV line to obtain
a black-hole mass of  \hbox{$M_{\rm BH} \sim 10^{11}
\mu_L^{-1}M_\odot$}}
%where a value of $\mu_L\sim4$ is obtained.
%In this work \hbox{$M_{\rm BH} \sim 10^{11} \mu_L^{-1}M_\odot$}
%are obtained using the linewith of the CIV line.}
is the lens magnification factor.} Equation~\ref{eq:vwi} could be
modified using a reliable SED describing the central source,
adding relativistic corrections, and calculating the force
multiplier at every point of the trajectory of the outflow. We
stress, however, that our simplified approach is sufficient to
provide first order approximations to the launching radius and the
time scales involved in the dynamics of the outflow. Equation
\ref{eq:vwi} can be written as $R_{\rm launch}/R_{S}= \Omega_f^2
(c/v_\infty)^2$; therefore for $\Omega_f\sim1.7$ $R_{\rm
launch}\sim3\times(c/v_\infty)^2 R_S$. The latter expression can
be used to obtain first order approximations of the launching
radius given the velocity of the outflow.

Assuming a radiation-driven wind, it is expected that the time
required to accelerate the outflow to fractions of $c$ is of the
order of 10$\frac{R_S}{c}$ (see Figure~\ref{fig:vrc}). For the
black-hole mass of \apm~of $M_{\rm BH}\approx 10^{10} M_{\odot}$
we estimate that the time to accelerate an absorber to
near-relativistic velocities is $\sim$weeks (rest-frame). We have
reported in this work probable variability of the high-energy
absorption features over a time-scale of $\sim$1 month
(rest-frame). This short time-scale variability is consistent with
the expected variability timescale of a radiation-driven wind.

\subsection{Constraints on Mass-Outflow Rate and Efficiency of the Outflow.} \label{s:out}

Based on our estimated values of the outflow velocities, column
densities, and launching radii we present constraints on the
mass-outflow rates and outflow efficiency associated with the
outflowing X-ray absorbers of \apm. The efficiency is defined as
the ratio of the rate of kinetic energy injected into the ISM and
IGM by the outflow to the quasar's bolometric luminosity, i.e.,
\begin{equation} \label{eq:kii}
\epsilon_K=\frac{1}{2}\frac{\dot{M}v^2}{L_{bol}}, ~{\rm where}~
\dot{M}=4 \pi R^2 \rho v f_c=4 \pi f_c \frac{R^2}{\Delta R} N_{\rm
H} m_p v,
\end{equation}
where $f_c$ is the covering fraction, $N_{\rm H}$ is the column
density, $R$ is the radius, and $\Delta R$ is the thickness of the
absorber. To estimate the efficiency we use the two
absorption-line model (APL+2AL; model 7 of Table~\ref{tab:modn}).
We calculate the bulk velocities of each outflow component based
on the energies of the absorption lines in model 7 and through the
use of equation (\ref{eq:Dop}) with $E_{\rm lab}$=6.7 keV and
$\theta=20^{\circ}$. As in \cite{Cha02,Cha03} we interpret the
high-energy absorption features as being due to highly ionized Fe
(\FeXXV\ K$\alpha$) in a gas with solar abundances, and we
estimate \lnh\ using a curve-of-growth analysis. In
Table~\ref{tab:epk} we present the outflow velocities, $v_{\rm
abs}$, the column densities, \lnh, the mass-outflow rates,
$\dot{M}$, and the outflow efficiencies, $\epsilon_K$, of the two
modeled absorbers of the outflow. We note that the values of the
column densities and velocities in Table~\ref{tab:epk} are
consistent with those found using the photoionization code XSTAR
(see Table~\ref{tab:xsta}).\footnote{The velocities obtained in
Table~\ref{tab:xsta} assume the redshifts  of the absorbers in
model XSTAR4 are due to the relativistic Doppler effect (see
equation~\ref{eq:Dop}).}

\begin{deluxetable*}
{cccccccccccc} \tabletypesize{\scriptsize}
%\tabletypesize{\footnootsize}
%\tabletypesize{\small}
\tablecolumns{12} \tablewidth{0pt} \tablecaption{Projected maximum outflow velocities, mass-outflow rates
and efficiencies of outflows in \apm\ ${}^{a}$.
\label{tab:epk}}

\tablehead{
\colhead{OBS} & \colhead{Instr.}  & \colhead{$v_{\rm
abs1}$} & \colhead{\lnh (abs1)} & \colhead{$\dot{M}$ (abs1)} &
\colhead{$\epsilon_K$ (abs1)} &  \colhead{$v_{\rm abs2}$} &
\colhead{\lnh (abs2)} &
\colhead{$\dot{M}$ (abs2)}  & \colhead{$\epsilon_K$ (abs2)}\\
& & [$c$] &  &  [$M_\odot \mu_L^{-1}{\rm yr}^{-1}$] &   & [$c$]
& & [$M_\odot \mu_L^{-1}{\rm yr}^{-1}$] &  }

\startdata

1 & XIS FI  & $0.20_{-0.02}^{+0.01}$ & 22.84$\pm$0.19 & $470^{+478}_{-319}$ & $0.02^{+0.02}_{-0.01}$ & $0.47_{-0.03}^{+0.02}$  &  23.07$\pm$0.27 & $2072^{+2388}_{-1496}$ & $0.5^{+0.6}_{-0.4}$\\

2 & XIS FI  & $0.27_{-0.02}^{+0.03}$ & 22.86$\pm$0.25 & $713^{+794}_{-507}$ & $0.05^{+0.06}_{-0.04}$ & $0.49_{-0.02}^{+0.02}$ &  22.81$\pm$0.39 & $1468^{+2071}_{-1167}$ & $0.4^{+0.5}_{-0.3}$\\

3 & XIS FI  & $0.17_{-0.01}^{+0.01}$ & 22.87$\pm$0.28 & $479^{+561}_{-350}$ & $0.01^{+0.02}_{-0.01}$ & $0.44_{-0.04}^{+0.04}$ &  23.05$\pm$0.19 & $1678^{+1707}_{-1141}$ & $0.3^{+0.4}_{-0.2}$\\

1 & XIS BI  & $0.17_{-0.02}^{+0.03}$ & 22.89$\pm$0.24 &  $475^{+520}_{-335}$ & $0.01^{+0.02}_{-0.01}$ & $0.46_{-0.03}^{+0.03}$ &  23.48$\pm$0.20 & $4772^{+4912}_{-3266}$ & $1.1^{+1.1}_{-0.7}$\\

2 & XIS BI  & $0.34_{-0.03}^{+0.02}$ & 23.01$\pm$0.36 & $1520^{+2040}_{-1179}$ & $0.19^{+0.25}_{-0.15}$ & $0.53_{-0.07}^{+0.05}$ &  23.09$\pm$0.36 & $2849^{+3822}_{-2210}$ & $0.8^{+1.1}_{-0.7}$ \\

3 & XIS BI  & $0.22_{-0.05}^{+0.04}$ &  23.10$\pm$0.21 & $962^{+1006}_{-663}$ & $0.05^{+0.05}_{-0.03}$ & $0.50_{-0.04}^{+0.04}$ &  23.14$\pm$0.24 & $2486^{+2722}_{-1753}$ & $0.7^{+0.7}_{-0.5}$ \\

Ch02 & ACIS BI & $0.20_{-0.01}^{+0.01}$ & 22.99$\pm$0.11 & $622^{+570}_{-420}$ & $0.03^{+0.02}_{-0.02}$ & $0.40_{-0.01}^{+0.01}$ &  23.07$\pm$0.12 &  $1507^{+1398}_{-979}$ & $0.3^{+0.2}_{-0.2}$  \\

Has02 & EPIC pn & $0.21_{-0.03}^{+0.03}$ &  23.10$\pm$0.36 & $1155^{+1550}_{-896}$ &  $0.10^{+0.14}_{-0.08}$ & $0.50_{-0.07}^{+0.06}$ & 23.36$\pm$0.30 &  $4500^{+5443}_{-3329}$ & $1.2^{+1.4}_{-0.9}$

\enddata

\tablenotetext{a}{The estimated values of the outflow properties
were based on fits that assumed an absorbed power-law model with
two absorption lines. The values of $\dot{M}$ and $\epsilon_K$ are
obtained by equation (\ref{eq:kii}) assuming $M_{\rm
BH}\sim 10^{12}\mu_L^{-1}M_\odot$ (see \S\ref{s:kin}) and $L_{\rm
bol}=7\times10^{15}\mu_L^{-1}L_\odot$ \citep{Irw98,Rei08}.}

\end{deluxetable*}

To obtain error bars for $\epsilon_K$ and $\dot{M}$ we performed a
Monte Carlo simulation, assuming a uniform distribution of the
parameters $f_c$, $R$ and $R/\Delta R$ around the expected values
of these parameters, and a normal distribution for \lnh\
(described by the parameters in Table~\ref{tab:epk}).
Specifically, we assume a covering factor lying in the range
$f_c=0.1-0.3$, based on the observed fraction of BAL quasars
\citep[e.g.,][]{Hew03} and a fraction $R$/$\Delta$$R$ ranging from
1 to 10 based on current theoretical models of quasar outflows
\citep[e.g.,][]{Pro00}. Note that in Table~\ref{tab:epk} we also
include the outflow parameters of the $\sim$90~ks
\chandra~observation of \apm\ performed in 2002 \citep{Cha02}.
Based on our estimated maximum outflow velocities ($v_{\rm max}
\sim 0.6c$) we expect that $R$ will be similar to $R_{\rm launch}$
and range between 3$R_S$ and 15$R_S$ (see Figure~\ref{fig:vrc}).
We note that this is a conservative assumption since larger values
of $R$ will result in larger mass-outflow rates and larger
efficiencies. Additionally, the short variability time-scales
($\sim$weeks) are also consistent with a launching radius of a few
times $R_S$. Variability in \apm\ over time-scales $\sim$weeks has
been previously reported (Chartas et al. 2003) based on the
differences of the high-energy absorption features detected in the
{\sl Chandra} and {\sl XMM-Newton} observations. Our current
analysis of the \suzaku\ observations of \apm\ allows us to
compare absorption features observed with the same instruments;
therefore, it avoids any possible systematic uncertainties due to
differences in the responses of the instruments.

Our results indicate that the average fraction of the total
bolometric luminosity of \apm\ injected into the IGM in the form
of kinetic energy is $\epsilon_K=0.7\pm0.3$. From the results of
Table~\ref{tab:epk} we also obtain the average mass-outflow rate
of \mbox{$\dot{M}=3324\pm915~\mu_L^{-1}M_{\odot}{\rm yr}^{-1}$}.
On the other hand the mass-accretion rate is $\dot{M}_{\rm
acc}=\frac{L_{\rm}}{\eta c^2}\sim 4840 \mu_L^{-1} M_{\odot}{\rm
yr}^{-1}$
%$\dot{M}_{\rm acc}=\frac{L_{\rm Edd}}{\eta c^2}$% \sim 400 M_{\odot}{\rm yr}^{-1}$
(assuming $\eta\approx0.1$ and $L_{\rm
bol}=7\times10^{15}\mu_L^{-1}L_\odot$). Therefore the mass-outflow
rate is comparable to the accretion rate. In the context of recent
models of structure formation \citep[e.g.,][]{Gra04,Spr05}, our
estimated values of $\epsilon_K$ in \apm~ suggest that these
outflows should be an important source of feedback in their host
galaxies and also play an important role in regulating the growth
of the central black hole.

To obtain an independent estimate of the mass-outflow rate
$\dot{M}$, we derive $nR^{2}$ based on the definition of
$\xi=L/(nR^2)$. Assuming $\lip\sim3.5$ (e.g., model XSTAR4,
Table~\ref{tab:xsta}) and an ionizing luminosity similar to the
X-ray luminosity $L_{\rm X}\sim 4\times10^{46}\mu_L^{-1} \lumin$,
we obtain $nR^2\sim 1\times10^{43}\mu_L^{-1}~{\rm cm}^{-1}$.
Therefore assuming an overall velocity of the outflow $\sim0.45c$
we find $\dot{M}\sim 9\times 10^3 \mu_L^{-1}M_\odot{\rm yr^{-1}}$.
The value of $\dot{M}$ derived from the best-fit ionization
parameter is comparable (within a factor of three) to the value
found for the second absorber abs2 in Table~\ref{tab:epk}, where
we have estimated the location of the absorber from variability
arguments. We caution, however, that the estimation of the
mass-outflow rate from the ionization parameter assumes a
spherical outflow illuminated by a point source. However, since we
expect that the X-ray absorber is located a few $r_{g}$ from the
X-ray source the point source approximation may not be accurate in
this case.

\section{CONCLUSIONS}

Our analysis of three long \suzaku\ observations of the BAL quasar
\apm\ indicates strong and broad absorption at \RF~energies of
$\lesssim$2~keV (low-energy) and \AEn\ (high-energy). Based on the
\Ft\ the low-energy absorption is significant at the $\gtrsim$99\%
and $\gtrsim$99.9\% levels in the front-illuminated (FI) and
back-illuminated (BI) \suzaku\ XIS spectra, respectively. The
high-energy absorption is  significant at $\gtrsim$99.9\% (FI
spectrum) and at $\gtrsim$99\% (BI spectrum) confidence,
respectively. The medium producing the low-energy absorption is a
nearly neutral absorber with a column density \lnh$\sim$23. The
medium producing the high-energy absorption appears to be
outflowing from the central source at near-relativistic velocities
and with large ionization parameters (\lipR), consistent with
results obtained from a previous {\sl Chandra} observation of this
object (Chartas et al. 2002). Simulations of highly ionized
near-relativistic winds performed by \cite{Sch07} indicate that
the resulting X-ray broad absorption profile may have the apparent
shape of an absorption edge, a notch, or a combination of
absorption lines depending on the assumed dynamics and degree of
ionization of the outflowing absorbers.

Our observations of the \AEn\ \RF\ features are well described by
a two component absorber model. We find that in epoch OBS1,
spectral fits with the two component \hbox{absorption-line} model
(APL + 2AL) are significantly better ($>$95\% confidence level)
than fits with \hbox{absorption-edge} models (APL + Ed). We note
that spectral fits with models that included ionized absorbers
with free iron abundances (models XSTAR2 and XSTAR3 in
Table~\ref{tab:xsta}) are consistent with no iron over-abundance
in all {\sl Suzaku} observations of \apm.

Our interpretation, of a near-relativistic outflowing absorbing
medium in a high ionization state ($2.75 \lesssim \lip \lesssim
4$), is consistent with our analysis of all the past \Rx\
observations of \apm. Our spectral analysis indicates that the
outflow velocities of the highly ionized absorbers detected in the
three \suzaku\ observations range between 0.1$c$ and 0.6$c$. The
maximum detected projected outflow velocity of $\sim$0.6$c$
constrains the angle between our line of sight and the wind
direction to be $\lesssim$36$^{\rm o}$. We find possible
variability of the high-energy absorption lines between epochs
OBS2 and OBS3 at the $\sim$99.9\% and $\sim$98\% significance
levels in the FI and BI spectra, respectively. Our spectral
analysis indicates that the variability is likely due to a change
in the outflow velocity of the absorber. The short time-scale
($\sim$month in the rest-frame) of this variability is probably
indicating that this absorber is strongly accelerated. This short
time-scale variability combined with the high ionization of the
absorbing material imply that the absorbers are launched from
distances $\lesssim$10$R_S$ from the central source.
\newline \indent Assuming our
interpretation that the absorption lines detected at \RF\ energies
of \AEn\ are due to \FeXXV, we estimate that a significant
fraction (0.7$\pm$0.3) of the total bolometric energy over the
quasar's lifetime is injected into the intergalactic medium of
\apm\ in the form of kinetic energy with a mass-outflow rate of
\mbox{$\dot{M}=3324\pm915~\mu_L^{-1}M_{\odot}{\rm yr}^{-1}$}.

\acknowledgments

We would like to thank Michael Eracleous and Konstantin Getman for
helpful discussions related to spectral models and statistical
tests implemented in this work. We acknowledge financial support
by NASA grant NNX08AZ67G and NASA LTSA grant NAG5-13035.

\end{document}